\documentclass[12pt]{article}
\usepackage[utf8]{inputenc}
\usepackage{graphicx,psfrag,epsf}
\usepackage{booktabs}
\usepackage{textgreek}
\usepackage{threeparttable}
\usepackage{float}
\usepackage{amsmath,amsfonts,amsthm,bm} 
\usepackage{url}
\usepackage{transparent}
\usepackage{adjustbox}
\usepackage{flafter}
\usepackage{lscape}
\usepackage{caption}
\usepackage{subcaption}
\usepackage{graphicx}
\usepackage{geometry}
\usepackage{footnote} 
\makesavenoteenv{tabular} 
\usepackage{verbatim}
\usepackage{natbib}
\usepackage{comment}
\usepackage{rotating}
\usepackage{hyperref}
\usepackage{mdframed}
\usepackage{lipsum}
\usepackage{array}
\newcolumntype{H}{>{\setbox0=\hbox\bgroup}c<{\egroup}@{}}
\usepackage{xcolor}
\usepackage{amsmath}
\usepackage{multirow}
\usepackage{bbm}
\usepackage{algorithmicx}
\usepackage{algorithm,algpseudocode}
\usepackage{longtable}

\usepackage{tikz}
\usetikzlibrary{matrix}
\usetikzlibrary{fit}
\usetikzlibrary{patterns}
\usetikzlibrary{calc}

\newcommand{\blind}{0}

\begin{document}

\def\spacingset#1{\renewcommand{\baselinestretch}%
{#1}\small\normalsize} \spacingset{1}

\def\spacingset#1{\renewcommand{\baselinestretch}%
{#1}\small\normalsize} \spacingset{1}


\if0\blind
{
  \title{\bf MIDAS-QR with 2-Dimensional Structure\thanks{
    The authors thank David Kohns, and Katalin Varga for discussions that shaped the methodology. Tibor Szendrei thanks the ESRC for PhD studentship as well as Heriot-Watt University for institutional support. The usual disclaimer applies.}}
  \author{Tibor Szendrei \footnote{Corresponding author: ts136@hw.ac.uk.}\\
    Department of Economics, Heriot-Watt University, UK. \\
    National Institute of Economic and Social Research, UK.\\
     \\
    Arnab Bhattacharjee\\
    Department of Economics, Heriot-Watt University, UK.\\
    National Institute of Economic and Social Research, UK.\\
    \\
    Mark E. Schaffer\\
    Department of Economics, Heriot-Watt University, UK.}
  \maketitle
} \fi

\if1\blind
{
  \bigskip
  \bigskip
  \bigskip
  \begin{center}
    {\LARGE\bf Title}
\end{center}
  \medskip
} \fi


\bigskip
\begin{abstract}
\noindent Mixed frequency data has been shown to improve the performance of growth-at-risk models in the literature. Most of the research has focused on imposing structure on the high-frequency lags when estimating MIDAS-QR models akin to what is done in mean models. However, only imposing structure on the lag-dimension can potentially induce quantile variation that would otherwise not be there.  In this paper we extend the framework by introducing structure on both the lag dimension and the quantile dimension. In this way we are able to shrink unnecessary quantile variation in the high-frequency variables. This leads to more gradual lag profiles in both dimensions compared to the MIDAS-QR and UMIDAS-QR. We show that this proposed method leads to further gains in nowcasting and forecasting on a pseudo-out-of-sample exercise on US data.
\end{abstract}


\noindent%
\noindent
\vfill

\spacingset{1.45} 


\section{Introduction}
Growth-at-risk (GaR), which is the measure of downside risk in the growth of GDP, has been a key concern for policy makers ever since the pivotal work of \citet{adrian2019vulnerable}. The idea of GaR is to look at GDP growth through the lens of Value-at-Risk, a statistic used in financial monitoring to quantify the possible financial losses within a certain period \citet{suarez2022growth}. In essence, GaR allows for modelling the nonlinear macro-financial linkages with the help of quantile regression of \citet{koenker1978regression}. In this way \citet{adrian2019vulnerable} shows how financial conditions have an significant impact on the lower tails of the GDP distribution.

Ever since the work of \citet{adrian2019vulnerable}, research has quickly built on these findings. \citet{kohns23hsbqr} find that there is no real difference in variability between upper and lower quantiles, but before crises episodes the forecasted distribution becomes multi-modal, a finding echoed by \citet{mitchell2022constructing} and \citet{adrian2021multimodality}. In relation to financial conditions driving nonlinearities in growth at risk, \citet{kohns2021decoupling} and \citet{szendrei2023revisiting} find evidence of quantile specific sparsity\footnote{Quantile specific sparsity is a situation where different variables are important for different parts of the distribution.}. \citet{loria2020understanding} find vulnerable growth behaviour using a regime switching model rather than quantile regression. 

Nowcasting GaR has also been an active area of research especially implementing quantile versions of the Mixed Data Sampling (MIDAS) method. \citet{ferrara2022high} applied a BMIDAS-QR (Bayesian MIDAS quantile regression) framework to nowcast Euro area GaR using two daily measures of financial conditions, optimally combining the forecasted quantiles to create GaR measures of the Euro Area. Their proposed method provides better nowcasting performance than the traditional BMIDAS as well as the BQAR(1) (Bayesian quantile AR(1) regression). Another recent nowcasting paper is \citet{xu2023mixed} who apply the MIDAS-QR approach to China, with constructed monthly factors of economic and financial activity in the Chinese economy. They find that the MIDAS-QR performs better than the traditional QR in both forecasting and nowcasting setups. \citet{carriero2020nowcasting} applied Bayesian quantile regression and Bayesian mixed frequency regression with stochastic volatility to a wide array of monthly and weekly variables. They find that more information enhances accuracy of nowcasting the tail of GDP growth. Also, Bayesian regression with stochastic volatility as well as Bayesian quantile regression performs better than simple quantile regression. Other studies such as \citet{ghysels2018quantile}, \citet{lima2020quantile} and \citet{xu2020mixed} highlight how high frequency data improve model performance in capturing higher moments of the density.

In the mixed frequency realm the focus has been on estimating quantiles independently. In this paper we propose extending the MIDAS-QR framework to jointly estimate the quantiles. By jointly estimating the quantiles we can impose structure on not just the coefficients across the lags (common in the MIDAS literaure) but also across the quantiles given the lag. To achieve this 2-dimensional structure we follow \citet{ferrara2022high}, \citet{lima2020quantile} and \citet{mogliani2021bayesian} in using the Almon lag polynomial structure to impose structure across the high frequency lags, and use adaptive non-crossing constraints of \citet{szendrei2023fused} to impose structure on the coefficients across the quantiles. Since adaptive non-crossing constraints can help recover different types of quantile estimators, we call this model the MIDAS-Generalised Non-Crossing Quantile Regression (MIDAS-GNCQR). To evaluate the performance of the model we consider a pseudo-out-of-sample US GaR application, using weekly NFCI as a financial conditions variable. Thus, we compare the proposed model with MIDAS-QR, combining an implementation of \citet{ferrara2022high} to US data, and with QR as proposed by \citet{adrian2019vulnerable}, which uses averaged NFCI values. We find that the MIDAS-GNCQR performs the best in both forecasting and nowcasting exercises. This highlights that for mixed-frequency QR, improvements can be gained by introducing some structure on the lag coefficients across the quantiles.

The paper is structured as follows. In section 2, first the MIDAS methodology is outlined, with particular focus on how the Almon polynomial can be used to introduce structure on the high frequency coefficients. Then the adaptive non-crossing constraints proposed by \citet{szendrei2023fused} are discussed and how these constraints enforce structure on the coefficients across quantiles. Section 3 briefly outlines the data used for estimation and transformations to create weekly versions of the NFCI corresponding to calendar months. Section 4 reports performance of the different models for select horizons. Here we deviate from the QR nowcasting literature in also discussing how the high-frequency coefficients compare between the different models. Finally, the paper concludes in section 5 with a discussion of potential further developments.

\section{Methodology}
The GaR framework of \citet{adrian2019vulnerable} relies on running a regression of GDP growth on past financial conditions (captured by some index) and past economic conditions (captured by lagged GDP). Using quantile regression allows the researcher to capture the nonlinear macro-financial linkages. \citet{ferrara2022high} and \citet{xu2023mixed} extend this approach by introducing higher frequency variables. Both papers utilise the MIDAS framework to tackle frequency mismatches. Formally the mixed frequency GaR can be represented as:

\begin{equation} \label{eq:GaR}
    y_{t+h} = x_t'\beta_q + w_{t}'\gamma_q + \epsilon_{t+h},
\end{equation}
where $x_t$ contains a constant and all the low frequency variables, and $w_t$ contains all the high-frequency variables, and $q$ is the quantile index of all the quantiles being estimated: $\tau=(\tau_1,\cdots,\tau_Q)$. In the above notation $w_t$ is set up in a MIDAS structure, i.e. for the case of a single high frequency variable it is an $1\times M$ vector where $M$ is the number of lags considered for the specific variable. We index a specific lag of the high frequency variable by $m$. Then, one can structure the data as $z_t=(x_t,w_t)$, and the coefficients as $\delta_q=(\beta_q,\gamma_q)$. The estimated coefficients $\hat{\delta}_q$ are obtained by minimising the tick-loss function of \citet{koenker1978regression}:

\begin{equation}\label{eq:QR}
\begin{split}
    \hat{\delta}&=\underset{\delta}{argmin}\sum^{Q}_{q=1}\sum^{T}_{t=1}\rho_q(y_t-z_t'\delta_{q})\\
    \rho_q(u)&=u(p-I(u<0)).
\end{split}
\end{equation}

Using the above coefficients one can calculate the predicted quantile by $\widehat{Q}_{y_{t+h}}(\tau_q|z)=\hat{\delta}_qz_{t}$. Note how one can easily obtain predicted quantiles for time $t+h$ by simply considering data until $t$. In this way we can create pseudo out-of-sample quantile estimates that can be used to evaluate the performance of the different estimators. Furthermore, if we consider estimating all the quantiles jointly, we can use forecast density evaluation metrics such as \citet{gneiting2011comparing}. In this way, we can evaluate what part of the density is better-estimated as more timely data are used in estimation.

At this point we deviate from other nowcasting papers in two ways. First, we estimate all quantiles jointly. This allows us to impose structure on the estimated coefficients across the quantiles. Second, we do not use the method of \citet{azzalini2003distributions} since it overwrites the fitted density, and it will then not be clear to what degree there are improvements in the fitted density.\footnote{Another reason to not fit a $t$-distribution using the estimated quantiles is that it purges multimodality from forecast densities. See \citet{kohns23hsbqr} and \citet{mitchell2022constructing} for evidence of multimodality in the forecasted distribution, particularly when the forecast was done right before crises episodes.}

Another key deviation is that we propose imposing a '2 dimensional' structure on the $\gamma$ coefficients. The first dimension of this structure is the $\gamma_q$ coefficients across the different lags $M$. This is achieved by not estimating the $\gamma_q$ coefficients directly, and instead approximating them by a polynomial function. In this way, we can trace the pattern of estimated $\gamma_q$ coefficients as we move across the $M$ lags. The smoothness of this curve can be regulated by changing the number of polynomials being estimated.

The second dimension along which structure is imposed is across the quantiles given a specific lag, i.e. the collection of the Q estimated $\gamma$ coefficients given a specific lag $m$. Specifically, we impose adaptive non-crossing constraints to regulate the variability of $\gamma_l$ across quantiles. Note that imposing this second structure is only possible because we jointly estimate the quantiles of interest.

\subsection{Structure along the lag dimension}

Depending on the frequency of the data, the size of the matrix $w$ can become large, which can pose problems in estimation. While it is true that if $M<<T$, equation (\ref{eq:GaR}) can be estimated without further amendments, for most economic time-series successive lags of the variable are likely to be highly correlated resulting in multicollinearity \citep{smith1976almon}. To address this issue, \citet{ghysels2004midas} proposed nonlinear least squares methods where, instead of directly estimating the $\gamma_q$ coefficients, one estimates polynomials in fractional lag orders. In this manner, some structure is imposed on the $\gamma_q$ coefficients. Formally, this implies the following equation:

\begin{equation}
    y_{t+h} = x_t'\beta_q + \sum^{M}_{m=1}\Big[\Tilde{B}(m;\theta_{q})L^{m/M}w_{t}'\Big]\gamma_q + \epsilon_{t+h},
\end{equation}
where $\Tilde{B}(c;\theta_{w,q})$ is a weighting function which depends on the parameters $\theta_{q}$ and lag order $m$. Note that in the above formulation, integer values of $h$ denote forecast horizons in the low frequency variable and $m/M$ denote fractional lags in the high frequency variables. While the weighting function is intentionally left as general as possible, there are various polynomial functions that can be used to approximate it, such as the Beta \citep{ghysels2016invest}, or the Almon \citep{lima2020quantile,ferrara2022high}. In this paper we implement the Almon lag polynomial structure to approximate the weighting function.

The Almon structure, originally proposed by \citet{almon1965distributed}, uses Weierstrass's theorem to approximate the lag structure of $\gamma_q$ by the use of polynomials. To implement the Almon lag structure we will use the so called ``direct method" of \citet{cooper1972two}, whereby the Almon lag polynomials are reparametrised as

\begin{equation} \label{eq:AlmonFullSpec}
    y_{t+h} = x_t'\beta_q + \Phi w_t'\theta_q + \epsilon_{t+h},
\end{equation}
where $\theta_q$ contains $p+1$ parameters, with $p<<M$ chosen by the practitioner, and $\Phi$ is a deterministic polynomial weighting matrix, with the $i^{th}$ row equal to $[0^i,1^i,\cdots,M^i]$ for $i=0,...,p$. In essence, the multicollinear high-frequency lag structure is approximated by the $(p+1)$ polynomial functions. This leads to a specification that is both linear and parsimonious.

An added advantage of the above specification is that further linear structure can be placed upon the value and slope of the lag polynomial. Following \citet{mogliani2021bayesian}, we impose restrictions such that the value and slope of the final high frequency lag parameter remains near zero and approaches it smoothly:
\begin{equation}
    \begin{split}
        &B(M;\theta_q)=0 \\
        &\nabla_M B(M;\theta_q)=0.
    \end{split}
\end{equation}
In this way the lag structure gradually `tails off' to zero. Note that by imposing these two restrictions (for each quantile), the number of parameters in equation (\ref{eq:AlmonFullSpec}) reduces from $(p+1)$ to $(p-1)$. One can impose further restrictions so long as $(p-r+1)>0$, where $r$ is the number of restrictions.

\subsection{Structure along the quantile dimension}

Together with the high-frequency lag dimension, we also place structure on the quantile dimension. To do so we impose adaptive non-crossing constraints from \citet{szendrei2023fused} based on the result that non-crossing constraints are a special type of fused shrinkage, with the hyperparameter being quantile specific. The key advantage of the method is that the non-crossing constraints can gradually be made tighter (or looser) to improves out-of-sample fit of the estimator. Because of the link between non-crossing constraints and fused shrinkage, such constraints regularise the amount of variation allowed between the different quantiles of $\gamma_l$.

For the remainder of the paper we define $\Tilde{z}_t=(x_t,\Phi w_t)$ and $\Tilde{\delta}_q=(\beta_q,\theta_q)$. Then, in the the estimation framework of equation (\ref{eq:QR}), instead of using the high-frequency lags directly, we use the converted Almon polynomials and estimate $\gamma_q$ through the Almon polynomial terms $\theta_q$.

Following the non-crossing quantile regression methodology of \citet{bondell2010noncrossing}, we set up the adaptive non-crossing constraints. First, we apply the min-max transformation to take all explanatory variables to the domain $[0,1]$. Next, we decompose the $\delta$ parameters of equation (\ref{eq:QR}) into differences: $(\upsilon_{0,1},...,\upsilon_{K,1})^T=\delta_{1}$ and $(\upsilon_{0,q},...,\upsilon_{K,q})^T=\delta_{q}-\delta_{q-1}$ for $q=2,...,Q$. We then follow linear programming notation and write the quantile differences as positive and negative components, i.e. $\upsilon_{j,q}=\upsilon^+_{j,q}-\upsilon^-_{j,q}$, where $\upsilon^+_{j,q}$ is its positive and $-\upsilon^-_{j,q}$ is its negative part. For each $\upsilon_{j,q}$ both components are non-negative and only one part is allowed to be non-zero. This leads to $Q-1$ constraints that can be added to equation (\ref{eq:QR}):

\begin{equation}\label{eq:BRW}
\begin{split}
    \hat{\delta}&=\underset{\delta}{argmin}\sum^{Q}_{q=1}\sum^{T}_{t=1}\rho_q(y_t-z_t'\delta_{q})\\
    &s.t.~\upsilon_{0,q}\geq \sum^K_{j=1}\upsilon^-_{j,q} ~ (q=2,...,Q) \\
    \rho_q(u)&=u(p-I(u<0)),
\end{split}
\end{equation}
where $K$ is the number of covariates in $z_t$, with the intercept being $z_{0,t}$.

Equation (\ref{eq:BRW}) leads to quantile estimates that do not cross in-sample. \citet{szendrei2023fused} build on this framework to introduce adaptive non-crossing constraints of the form: 
\begin{equation}\label{eq:AdaNC}
\begin{split}
    \hat{\delta}&=\underset{\Tilde{\delta}}{argmin}\sum^{Q}_{q=1}\sum^{T}_{t=1}\rho_q(y_t-\Tilde{z}_t'\Tilde{\delta}_{q})\\
    &s.t.~\upsilon_{0,q}+\sum^K_{j=1} \Big[ \Bar{Z_j} - \alpha(\Bar{Z_j} - min(Z_j)) \Big]\upsilon_{j,q}^+\geq \sum^K_{j=1} \Big[\Bar{Z_j} + \alpha(max(Z_j)-\Bar{Z_j}) \Big] \upsilon_{j,q}^-\\
    & (q=2,...,Q) \\
    \rho_q(u)&=u(p-I(u<0)).
\end{split}
\end{equation}
By simply adjusting the $\alpha$ parameter one can tighten or loosen the non-crossing constraints, which in turn regulates the size of allowed quantile variation in the coefficients.

Equation (\ref{eq:AdaNC}) imposes a 2 dimensional structure on the high frequency lag coefficients. First, the lags of the high frequency variable are approximated as $\Tilde{z}_t$ which includes the Almon polynomials rather than the raw high frequency data. This approximates the value of $\gamma_l$ by the $\theta_q$ terms. Second, we allow the $\theta_q$ terms to vary with each quantile, but restrict the amount of variation through adaptive non-crossing constraints. In this way, we allow lag profiles that are quantile specific, but not too dissimilar from each other. Through the use of adaptive non-crossing constraints, these quantile specific lag profiles are only allowed to deviate from each other if it leads to better out-of-sample fit. Furthermore, if $\alpha \geq 1$ the estimated quantiles are guaranteed to not cross in-sample.

\begin{figure}[t]
    \centering
\includegraphics[width=0.9\textwidth]{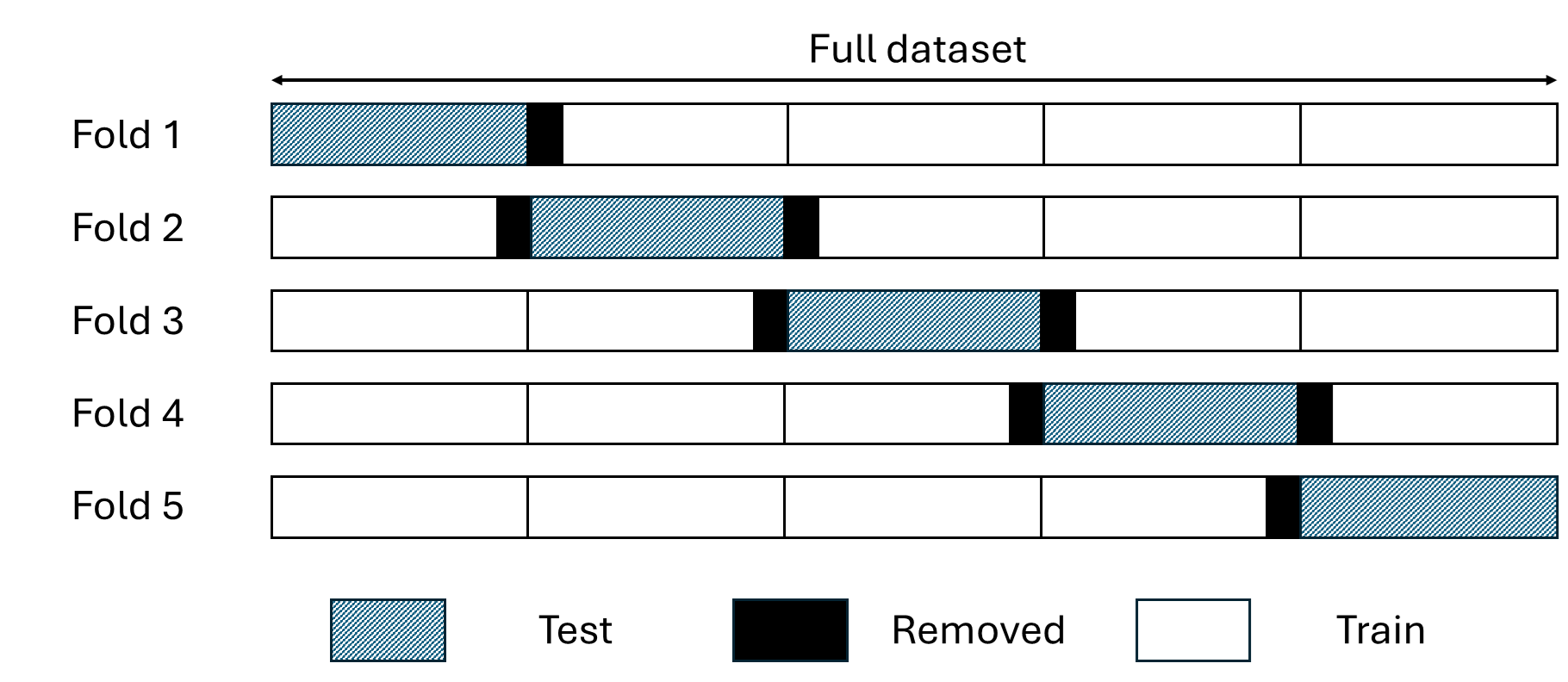}
\caption{hv-Cross Validation}
\label{fig:CV-fig}
\end{figure}

To find the optimal $\alpha$ parameter, we follow \citet{szendrei2023fused} and use cross validation (CV). In particular we use the $hv$-block CV setup of \citet{racine2000consistent}. This block setup is a time series version of the popular $k$-fold cross validation. The ``$h$" in the $hv$-block is the forecast horizon, and is equal the data around the testing sample that we do not use for estimation. When we are conducting nowcasts (i.e. situations where the forecast horizon is larger than 0 but less then 1) we simply set $h$ to 1. For all horizons we conduct 10 fold cross validation to find the $\alpha$ parameter. See \citet{cerqueira2020evaluating} for further details on the performance of the $hv$-block CV and other CV methods for time series.

Figure (\ref{fig:CV-fig}) visualises this block setup and which data are removed from the dataset. The reason to remove the data points around the test datasets is to ensure that no data leakage occurs when evaluating the performance of the model. When data are dependent, the information from the test dataset can leak into the training set. This can inadvertently lead to overfitting and hence poor generalization. Removing the data around the test dataset can guard against this occuring.

\section{Data and Evaluation Criteria}
To analyse the US GaR with mixed frequency data, we follow \citet{adrian2019vulnerable} and focus on fitting annualised GDP growth with its own lag, and the NFCI, which is an index measuring the financial conditions in the economy. However, unlike the original GaR, we use weekly NFCI values in the estimation. Furthermore, similar to \citet{ferrara2022high}, we also include a higher frequency economic activity variable: the index of industrial production (IP).\footnote{ \citet{ferrara2022high} include high frequency data on Purchasing Managers Index.} We use 12 lags for the weekly NFCI, and 6 lags for the monthly IP variables. Similar to \citet{ferrara2022high}, we set $p=3$ to approximate the high-frequency lag profiles. Then, the final model estimated takes the form:
\begin{equation}
\begin{split}
    y_{t+h} =& \beta_{0,q} + y_t'\beta_q + \sum^{12}_{m=1}\Big[\Tilde{B}(m;\theta_{NFCI,q})L^{m/12}NFCI_{t}'\Big]\gamma_{NFCI,q} \\
    &+ \sum^{6}_{m=1}\Big[\Tilde{B}(m;\theta_{IP,q})L^{m/6}IP_{t}'\Big]\gamma_{IP,q} + \epsilon_{t+h}
\end{split}
\end{equation}

Note, however, that the weekly NFCI variable is not reported by calendar week. Due to this, if one would use the raw weekly NFCI variable, there will be some months with 5 or 6 weeks. To amend this we first convert the raw weekly NFCI variables using the method outlined in \citet{smith2016google}. The key point in the conversion is that it creates months with equal numbers of weeks. To generate these new weekly variables, first we create daily NFCI values by making repeat copies of the weekly values of NFCI. Then we take the calendar weekly averages: Week 1 is the average of the values between the 1st and the 7th of the month, Week 2 is the average of the dates 8th to 14th, Week 3 the average of 15th to 21st, and we consider Week 4 data as the average of the 22nd until the last day of the month. Admittedly, week 4 is generally a little longer than weeks 1-3, but the advantage of this approach is that 12 lags of weekly NFCI always cover a quarter worth of data.

To evaluate the pseudo-out-of-sample fits of the different estimators, the quantile weighted CRPS (qwCRPS) of \citet{gneiting2011comparing} is chosen as a scoring rule. This measure uses the Quantile Score (QS), which is the weighted residual of an observation, $\hat{y}_{t+h,p}$, and assigns different weights to the different quantiles' QS. Formally the qwCRPS is calculated as:

\begin{equation}
    qwCRPS_{t+h} = \int^1_0 \; \omega_q QS_{t+h,q}dq,
\end{equation}
where $\omega_q$ denotes a weight assigned to the different quantiles which allows us to evaluate the different parts of the fitted density. We consider 4 weighting schemes: $w_q^1=\frac{1}{Q}$ places equal weight on all quantiles, which is equivalent to taking the average of the weighted residuals; $w_q^2=q(1-q)$ places more weight on central quantiles; $w_q^3=(1-q)^2$ places more weight on the left tail; and finally $w_q^4=q^2$ places more weight on the right tail.

For model evaluation, we consider 3 different types of models. First, the proposed MIDAS method with 2-dimensional structure (MIDAS-GNCQR), and the MIDAS with only 1-dimensional structure (MIDAS-QR), which is the estimator proposed in \citet{ferrara2022high} and \citet{xu2023mixed} are estimated. To compare the impact of the different structures on the high frequency lag coefficients, a MIDAS with no structure (UMIDAS-QR) is also estimated, but the fits of this model are not evaluated. To show the forecast improvements of the different structures, a QR model is also estimated, using the quarterly averages of NFCI and IP as the data. As such, we compare 3 models, but the reference model will be different depending on whether we are interested in the lag coefficients or the forecast performance.

We fit a nowcast model for 11 of the 12 weeks in a quarter, as well as a forecast model with $h=1$ and $h=4$. To analyse the impact of the imposed structure on the coefficients we focus on $h=4$, and $h=0.08$ (which is 1 week away from the end of the quarter). Detailed statistics for the other horizons are reported in the appendix. For all estimation periods we estimate 11 quantiles: every 10th quantile, as well as the 25th and 75th quantile.

Because of the COVID-19 period being part of our sample, comparing predictive accuracy becomes difficult, since it has a large influence on the variation of the predictive errors. As such, we also show the pre-COVID predictive performance of the estimators. To test the statistical significance between the estimators, we utilise \citet{diebold1995comparing} test.

\section{US MIDAS-GaR}
\subsection{Forecast performance}
\subsubsection{Coefficient profiles}

\begin{figure}
    \centering
    \includegraphics[width=\textwidth]{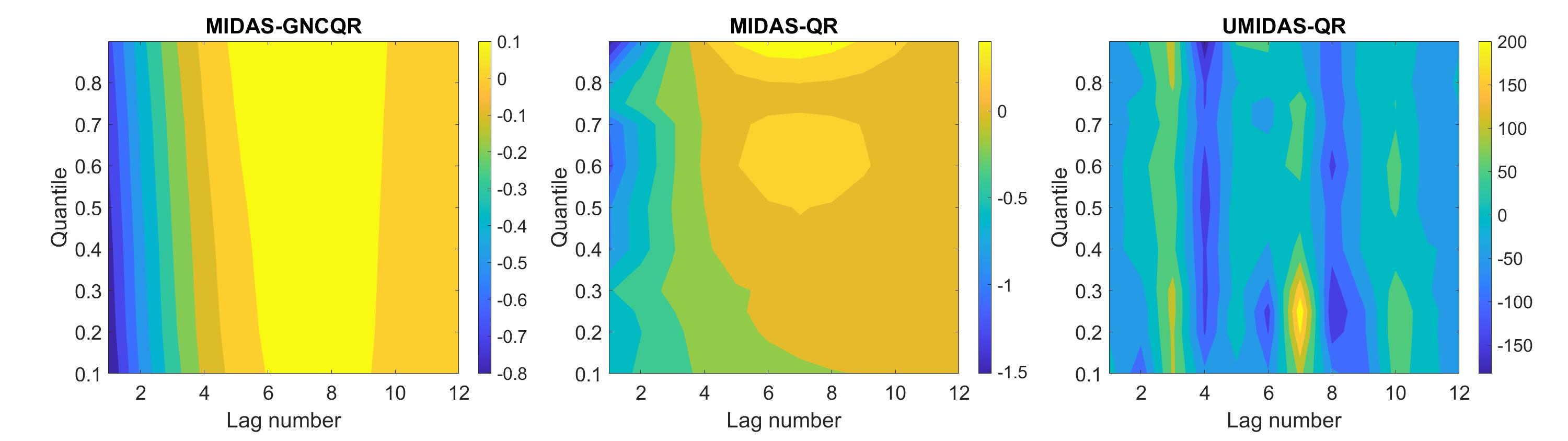}
    \caption{Effects of NFCI for $h=4$}
    \label{fig:2DCoeffNFCI_h4}
\end{figure}

\begin{figure}
    \centering
    \includegraphics[width=\textwidth]{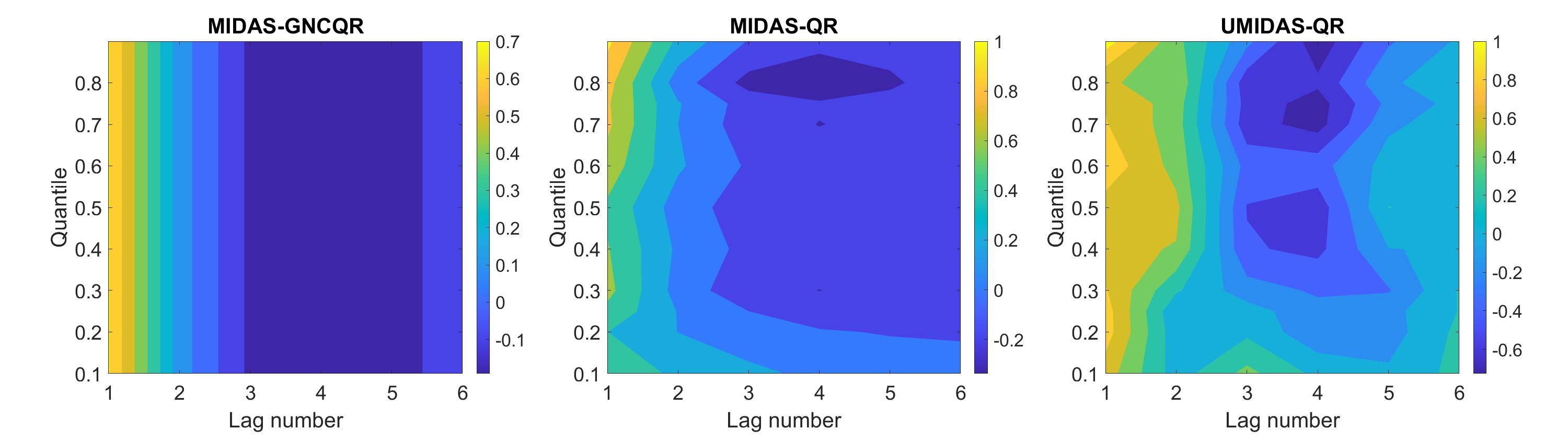}
    \caption{Effects of IP for $h=4$}
    \label{fig:2DCoeffIP_h4}
\end{figure}

We show effects across the 2-dimensional structure using contour plots. Specifically, figures (\ref{fig:2DCoeffNFCI_h4}) and (\ref{fig:2DCoeffIP_h4}) plot the effects by high-frequency lags for the different quantiles for $h=4$ for NFCI and IP respectively. The quantiles are shown on the $y$-axis, the high frequency lag orders on the $x$-axis, and the colour in the figures represent estimated coefficient values of $\gamma_{q,m}$. The right figure is the UMIDAS-QR which has no structure imposed on either dimension. The middle figure is the MIDAS-QR which has structure imposed on the lags through a polynomial approximation (and end-point restrictions); this is equivalent to imposing structure on the coefficients as it moves on the $x$-axis. The left figure is the MIDAS-GNCQR which imposes structure on both the lags and the quantiles, i.e. 2-dimensional structure.

Clearly, the structure has an impact on the estimated scale of the coefficients, with UMIDAS-QR showing the largest variation in potential coefficient values. This is particularly extreme for the NFCI in figure (\ref{fig:2DCoeffNFCI_h4}), with the values of $\gamma_{q,l}$ ranging between $-200$ to $200$. This wide range is likely because of high collinearity among the lags of the NFCI. Comparing the MIDAS-QR and the UMIDAS-QR shows how using the Almon polynomial setup, to ensure a smooth coefficient profile along the nowcast lag dimension, alleviates much of the problem.\footnote{Another solution to tackling the collinarity in UMIDAS is using regularisation techniques, such as ridge regression, on the nowcast lag coefficients.} 

The overall quantile profiles change drastically when we introduce the structure on the lags. In particular, we can see that for the UMIDAS-QR there is far more variation in the lag dimension than in the quantile dimension while for the MIDAS-QR we see the opposite. This justifies the introduction of constraints that regulate the quantile dimension. We can see in figure (\ref{fig:2DCoeffIP_h4}) that for the IP lag coefficients, most of the quantile variation is purged, and there is only variation across the lags.


Considering the NFCI lag structure in figure (\ref{fig:2DCoeffNFCI_h4}), we can see that imposing structure on both the lag and quantile dimensions leads to a coefficient plot that is a mix of the UMIDAS-QR and the MIDAS-QR. In particular, the scale of the coefficients is not as extreme as the UMIDAS-QR, but it also does not shift most of the variation into the quantile dimension. Instead, for the MIDAS-GNCQR we see a more gradual change in the quantile profile as we move across the lags. Both in the cases of NFCI and IP, the proposed 2-dimensional structure achieves good regularisation and renders quantile profiles smoothly varying, realistic and interpretable.

\begin{figure}
     \centering
     \begin{subfigure}[b]{0.49\textwidth}
         \centering
         \includegraphics[width=\textwidth]{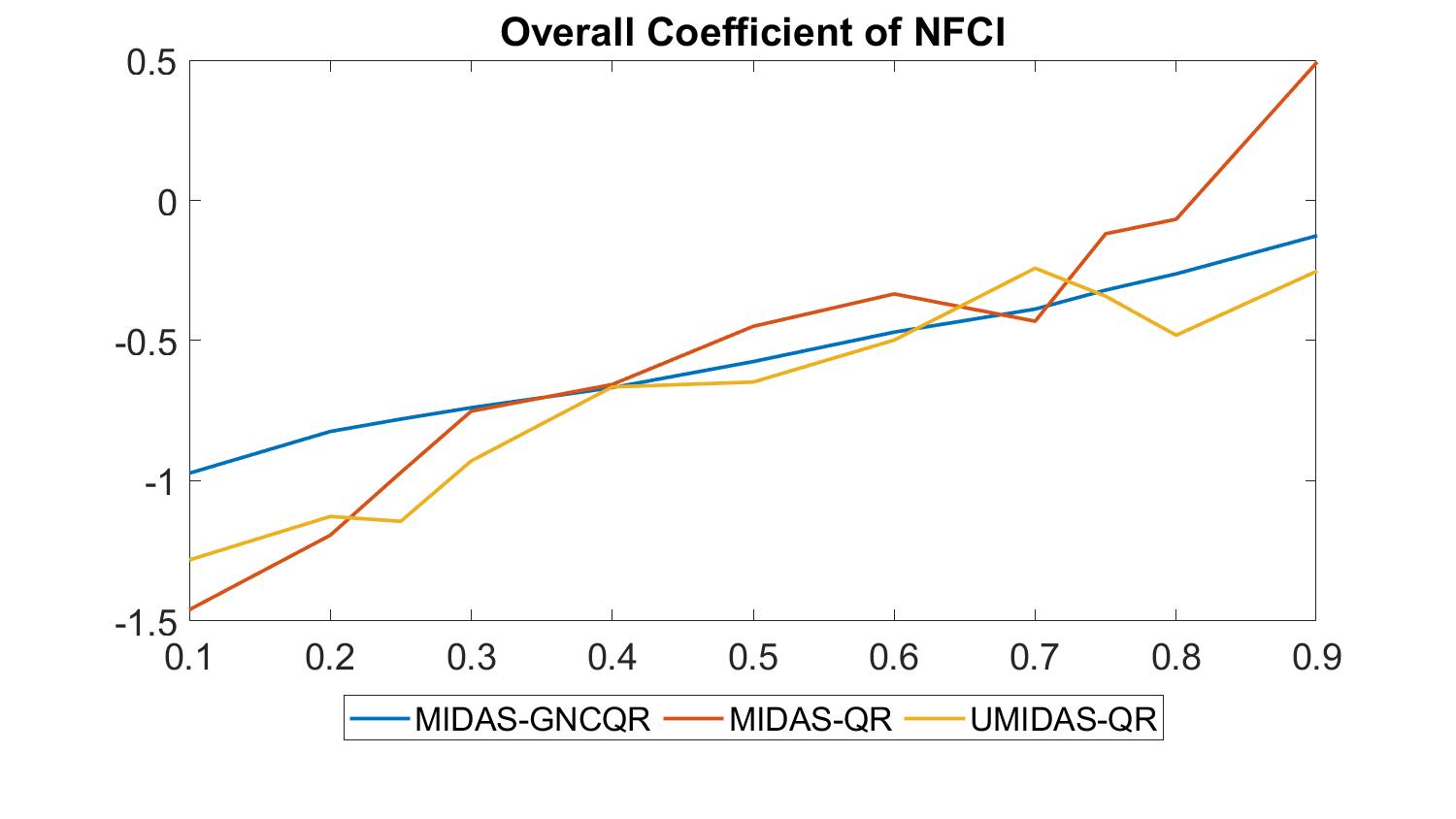}
         \caption{NFCI overall effect for $h=4$}
         \label{fig:NFCI_h4}
     \end{subfigure}
     \hfill
     \begin{subfigure}[b]{0.49\textwidth}
         \centering
         \includegraphics[width=\textwidth]{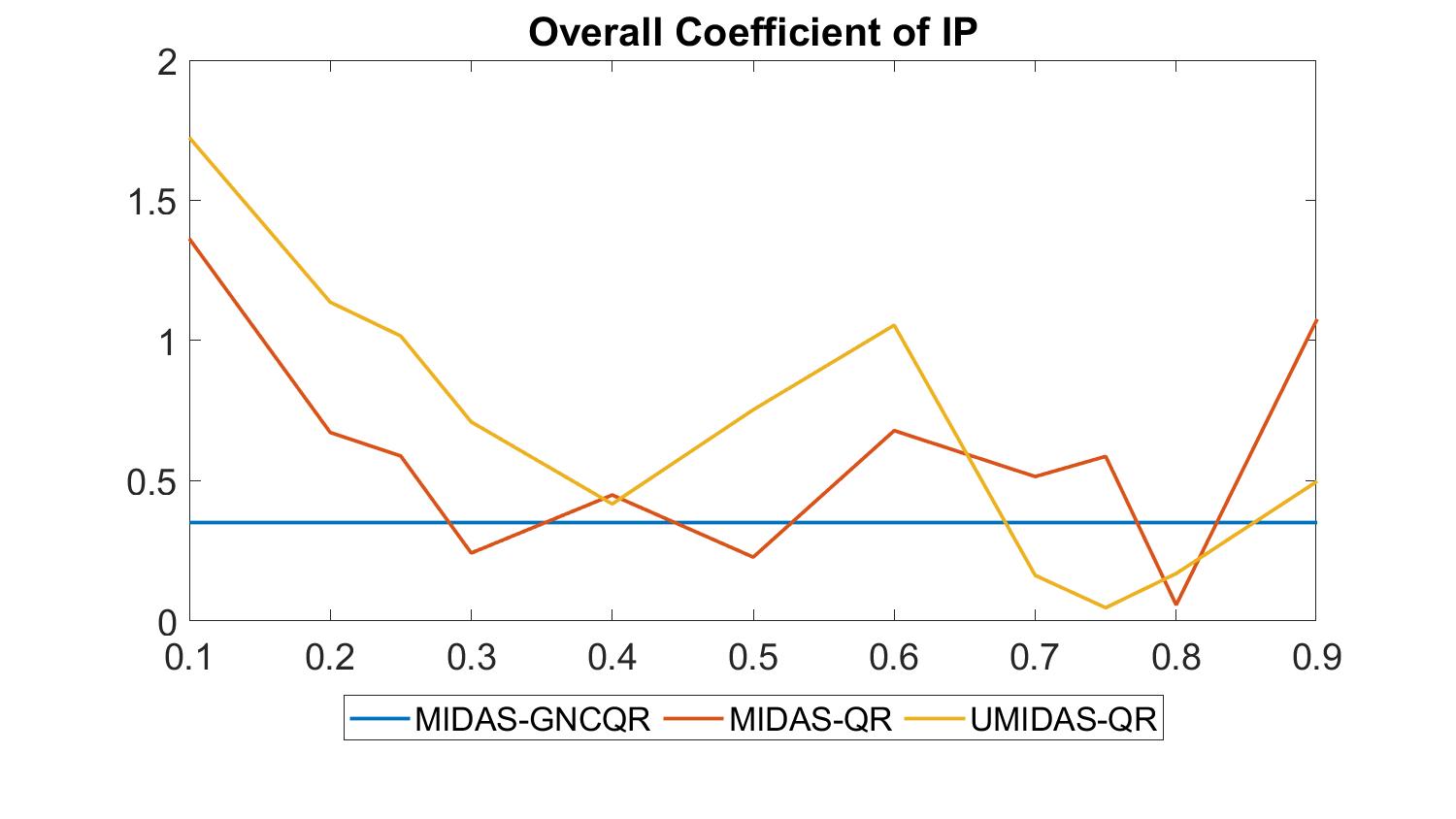}
         \caption{IP overall effect for $h=4$}
         \label{fig:IP_h4}
     \end{subfigure}
     \caption{Overall coefficients for $h=4$}
\end{figure}

We also show the overall effect of the high-frequency variable in figures (\ref{fig:NFCI_h4}) and (\ref{fig:IP_h4}) for NFCI and IP respectively. These effects are calculated by $\Phi\gamma_q\iota$, where $\iota$ is a vector of 1's with dimension as the number of high-frequency lags for the given variable. We can see that MIDAS-GNCQR shows no quantile variation for IP, which is in line with \citet{adrian2019vulnerable} who find no significant quantile variation for GDP. The overall effect of the NFCI is a negative, but increasing, coefficient profile for all estimators. Importantly, the UMIDAS-QR and MIDAS-GNCQR show that the overall impact of the NFCI is always negative, while the MIDAS-QR shows positive effects of NFCI at the upper tail. \citet{adrian2019vulnerable} argues that financial conditions are more likely to have a significant negative impact if any at all. This further highlights how imposing structure on both dimensions leads to an estimator that effectively combines the features of both UMIDAS-QR and MIDAS-QR.

\begin{table}
\centering
\caption{Forecast Results}
\label{tab:ForcRes}
\resizebox{\columnwidth}{!}{%
\begin{tabular}{lr|cccc|cccc}
\hline
& & \multicolumn{4}{c|}{Full Sample} & \multicolumn{4}{c}{Pre-COVID}\\
 &  & $\omega_1$ & $\omega_2$ & $\omega_3$ & $\omega_4$& $\omega_1$ & $\omega_2$ & $\omega_3$ & $\omega_4$ \\ \hline \hline
\multicolumn{2}{l|}{$h=1$} &  &  &  &  \\
 & MIDAS-GNCQR & 0.911 & 0.175 & 0.280 & 0.456 & 0.638 & 0.125 & 0.192 & 0.322 \\
 & MIDAS-QR & 0.926 & 0.177 & 0.284 & 0.465 & 0.646 & 0.126 & 0.195 & 0.325 \\
 & QR & 1.041 & 0.198 & 0.319 & 0.524 & 0.671 & 0.131 & 0.206 & 0.334 \\ \hline
\multicolumn{2}{l|}{$h=4$} &  &  &  &  \\
 & MIDAS-GNCQR & 0.616 & 0.120 & 0.180 & 0.316 & 0.495 & 0.098 & 0.145 & 0.252 \\
 & MIDAS-QR & 0.629 & 0.122 & 0.183 & 0.323 & 0.502 & 0.099 & 0.150 & 0.252 \\
 & QR & 0.649 & 0.126 & 0.189 & 0.333 & {0.524}* & 0.103 & {0.157}* & 0.263 \\ \hline
\end{tabular}
}%

{\raggedright \footnotesize Note: Statistically significant differences at the 10\% (*), 5\% (**), and 1\% (*) level are shown. In all cases, MIDAS-GNCQR is the reference estimator for the \citet{diebold1995comparing} test.\par}
\end{table}

\subsubsection{Evaluation metrics}

The forecast evaluation metrics are shown in table (\ref{tab:ForcRes}). First, we can see that MIDAS helps in forecast performance: both the MIDAS-QR and the MIDAS-GNCQR improves forecast performance of the GaR for both forecast horizons. Furthermore, we can see that imposing structure on both the lag dimension and quantile dimension leads to further improvements over the MIDAS-QR. This improvement is across all parts of the distribution as the MIDAS-GNCQR yields the best forecast performance regardless of the weighting scheme used. We note, that these difference on the full sample are not significant. However, the MIDAS-GNCQR does provide significantly better performance on the pre-COVID sample for $h=4$. The lack of significant differences in the full sample are likely on account of the large variation in GDP data during the COVID-19 period. The fitted models have difficulty picking up these extreme movements which results in large forecast errors. Importantly, these large errors have a large influence on the forecast scores, which in turn leads to wider forecast scores distributions. Nevertheless, the findings highlight the MIDAS-GNCQR's ability to regularise quantile and lag profiles leads to better model fits. 

\begin{figure}[]
    \centering
    \includegraphics[width=\textwidth]{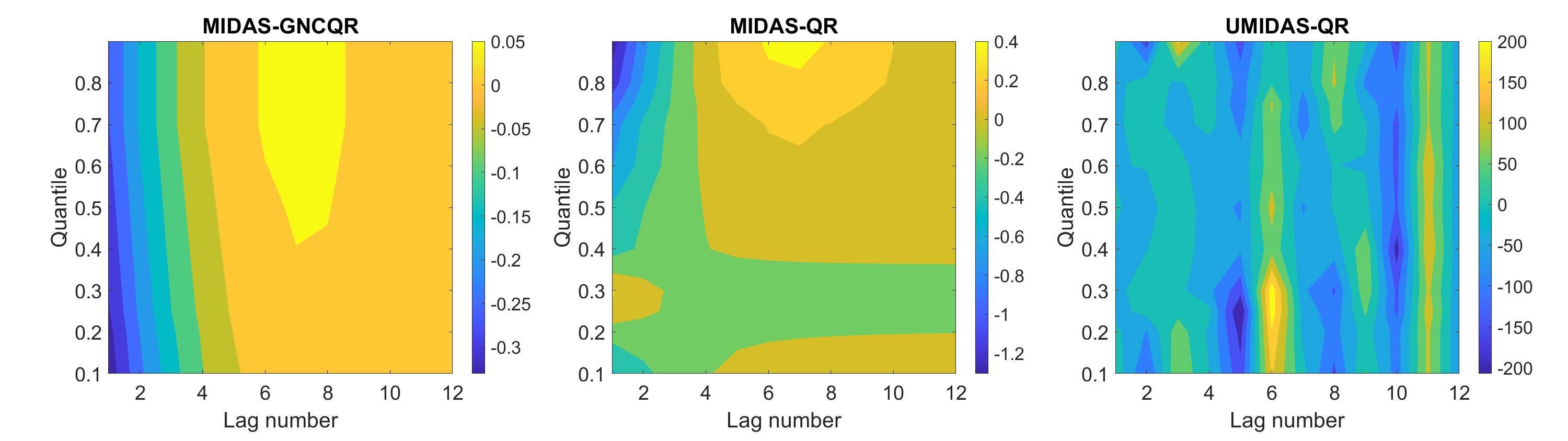}
    \caption{Effects of NFCI for $h=0.08$}
    \label{fig:2DCoeffNFCI_h0.08}
\end{figure}

\begin{figure}
    \centering
    \includegraphics[width=\textwidth]{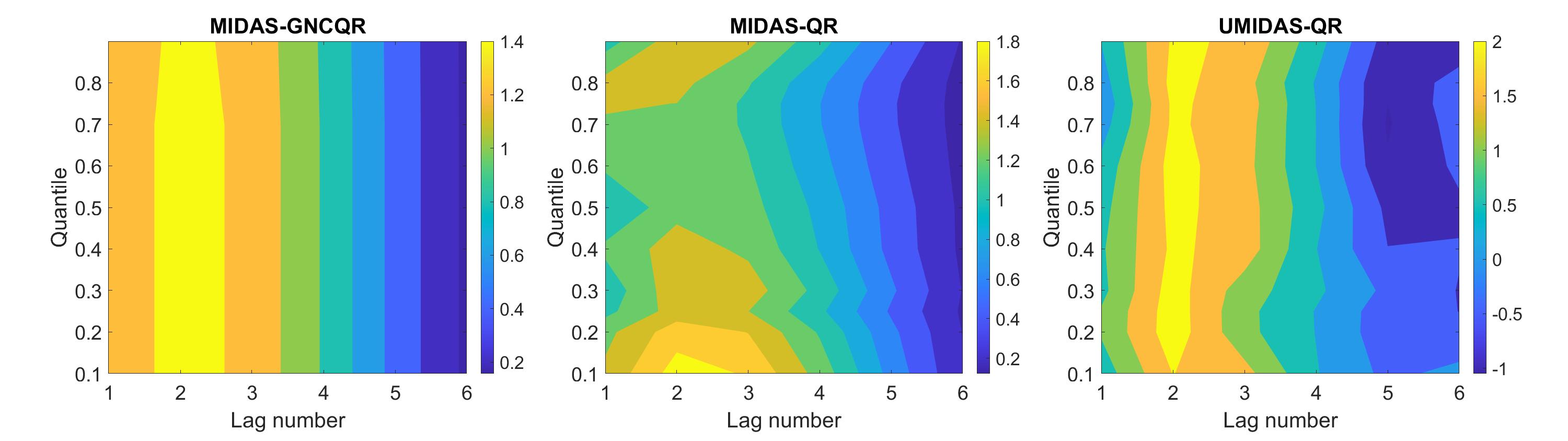}
    \caption{Effects of IP for $h=0.08$}
    \label{fig:2DCoeffIP_h0.08}
\end{figure}

\begin{figure}
     \centering
     \begin{subfigure}[b]{0.49\textwidth}
         \centering
         \includegraphics[width=\textwidth]{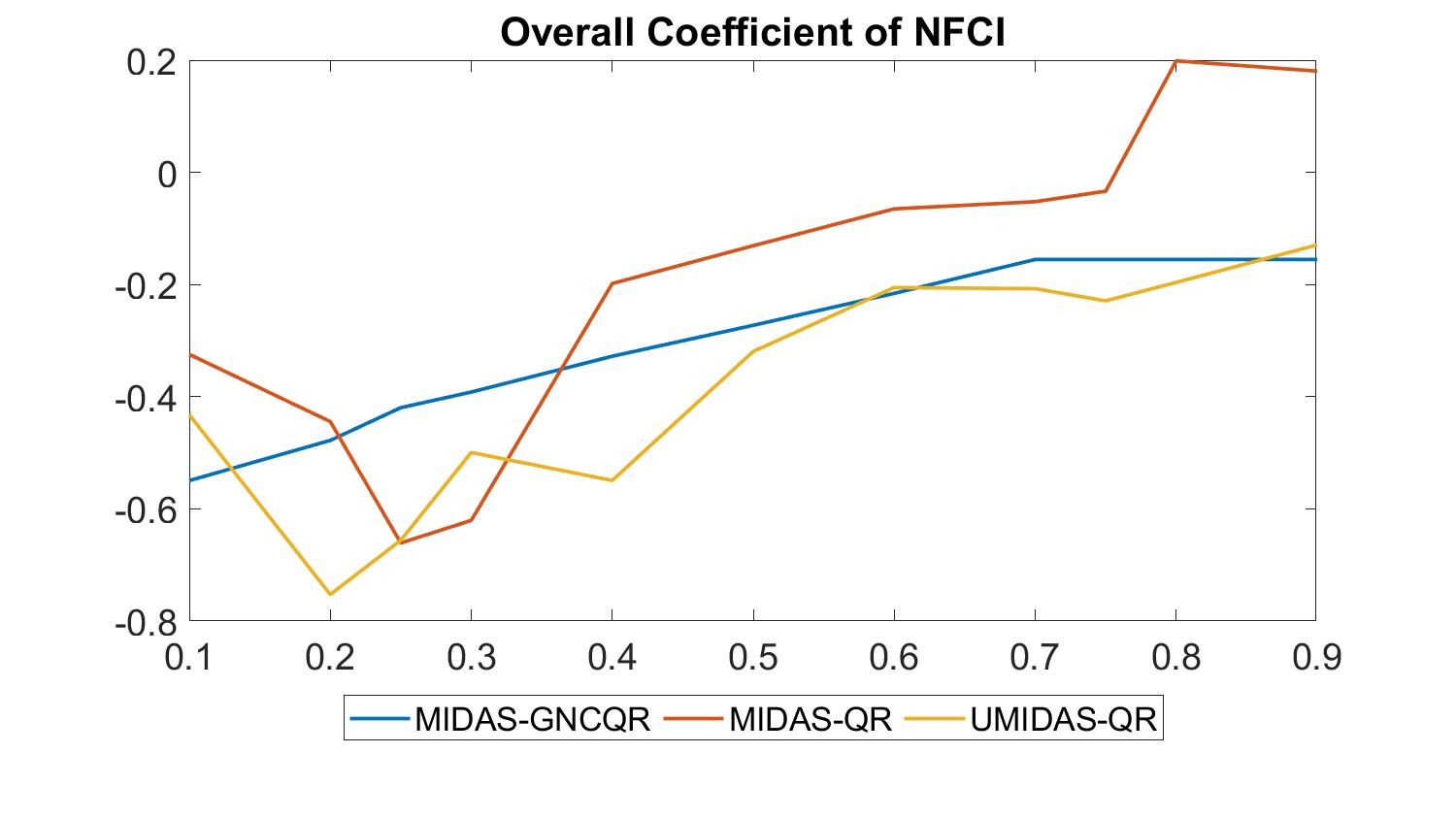}
         \caption{NFCI overall effect for $h=0.08$}
         \label{fig:NFCI_h0.08}
     \end{subfigure}
     \hfill
     \begin{subfigure}[b]{0.49\textwidth}
         \centering
         \includegraphics[width=\textwidth]{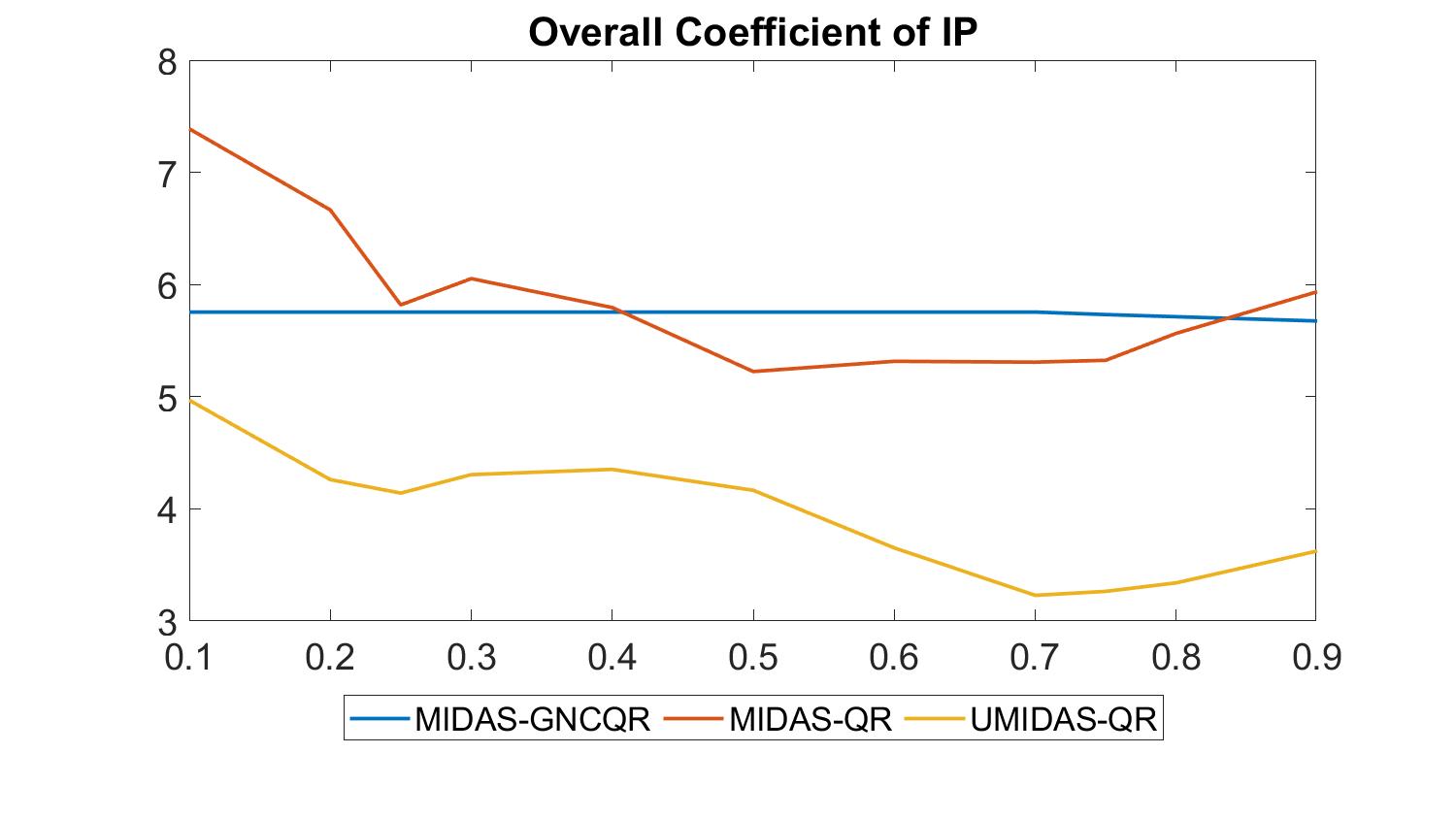}
         \caption{IP overall effect for $h=0.08$}
         \label{fig:IP_h0.08}
     \end{subfigure}
     \caption{Overall coefficients for $h=0.08$}
\end{figure}

\newpage
\subsection{Nowcast performance}
\subsubsection{Coefficient profiles}

Now, we evaluate nowcasts, first discussing the coefficient profiles of the estimators. These profiles are presented in figures (\ref{fig:2DCoeffNFCI_h0.08}) and (\ref{fig:2DCoeffIP_h0.08}) for the NFCI and IP respectively. Like the forecasting case, the range of coefficient estimates for UMIDAS-QR is extremely wide. While the MIDAS-QR remedies this, it leads to more variation in the quantile dimension. The MIDAS-GNCQR yields a more gradual change across both lag and quantile dimension, leading to an overall smoother surface.

The overall effects are presented in figures (\ref{fig:NFCI_h0.08}) and (\ref{fig:IP_h0.08}) for NFCI and IP respectively. Mirroring the forecasting case, the MIDAS-GNCQR shows no quantile variation for the IP, and negative but increasing profile for the NFCI. Similar to the forecasting case, the MIDAS-GNCQR yields an NFCI quantile profile that remains negative across all quantiles. However, compared to the forecasting case, the estimated coefficients for the IP are larger for all estimators, and the range of values for the NFCI is smaller.

\subsubsection{Evaluation metrics}

The nowcast evaluation results are presented in table (\ref{tab:NowRes}). We only present a few nowcast horizons, namely every 4th week, starting from the smallest $h$. This means that we report performance for the 1/12, 5/12, and 9/12 nowcast horizons. We note that this is a pseudo-nowcast exercise, since release calendars are not factored in when making the estimates. For this nowcast exercise we only compare the MIDAS-QR with the MIDAS-GNCQR.

\begin{table}[]
\centering
\caption{Select Nowcast Results}
\label{tab:NowRes}
\resizebox{\columnwidth}{!}{%
\begin{tabular}{lr|cccc|cccc}
\hline
& & \multicolumn{4}{c|}{Full Sample} & \multicolumn{4}{c}{Pre-COVID}\\
 &  & $\omega_1$ & $\omega_2$ & $\omega_3$ & $\omega_4$& $\omega_1$ & $\omega_2$ & $\omega_3$ & $\omega_4$ \\ \hline \hline
 \multicolumn{2}{l|}{$h=0.08$} &  &  &  &  \\
 & MIDAS-GNCQR & 0.815 & 0.158 & 0.243 & 0.414 & 0.569 & 0.112 & 0.169 & 0.288\\
 & MIDAS-QR & 0.828 & 0.160 & 0.245 & 0.423 & {0.572}* & {0.113}* & {0.172}* & 0.287\\ \hline
\multicolumn{2}{l|}{$h=0.42$} &  &  &  &  \\
 & MIDAS-GNCQR & 0.753 & 0.147 & 0.225 & 0.381 & 0.625 & 0.123 & 0.192 & 0.311\\
 & MIDAS-QR & 0.762 & 0.148 & 0.230 & 0.383 & 0.632 & 0.124 & 0.196 & 0.311\\ \hline
\multicolumn{2}{l|}{$h=0.75$} &  &  &  &  \\
 & MIDAS-GNCQR & 0.916 & 0.176 & 0.282 & 0.459 & 0.639 & 0.125 & 0.192 & 0.323\\
 & MIDAS-QR & 0.936 & 0.179 & 0.289 & 0.468 & {0.648}* & {0.127}* & {0.196}* & 0.325\\ \hline
\end{tabular}
}%

{\raggedright \footnotesize Note: Statistically significant differences at the 10\% (*), 5\% (**), and 1\% (*) level are shown. In all cases, MIDAS-GNCQR is the reference estimator for the \citet{diebold1995comparing} test.\par}
\end{table}

The results of table (\ref{tab:NowRes}) are in line with the findings of table (\ref{tab:ForcRes}), namely that the MIDAS-GNCQR provides further gains over the MIDAS-QR. Furthermore, mirroring the forecasting case, the MIDAS-GNCQR provides gains over all parts of the distribution as highlighted by lower weighted quantile score for all weighting schemes for each forecast horizons considered.\footnote{We report the results for all nowcast horizons in the appendix. The overall better forecast performance is true for all horizons except $h=2/12$ where the MIDAS-QR is marginally better at the right tail.} Just like in the forecast setting, we also see that the COVID period observations have a large influence on the measures, since none of the differences seem to be statistically significant. This changes when looking at the pre-COVID sample, where the MIDAS-GNCQR provides significantly better nowcast results for 2 of the 3 horizons. Furthermore, 3 out of 4 measures point towards the MIDAS-GNCQR providing significantly better nowcast results. These results highlight that introducing structure on not only the lag of the high-frequency variable but on the quantile structure as well lead to a better nowcasting performance.

\subsection{Fitted Densities before the 2008 crisis}

\begin{figure}
    \centering
    \includegraphics[width=\textwidth]{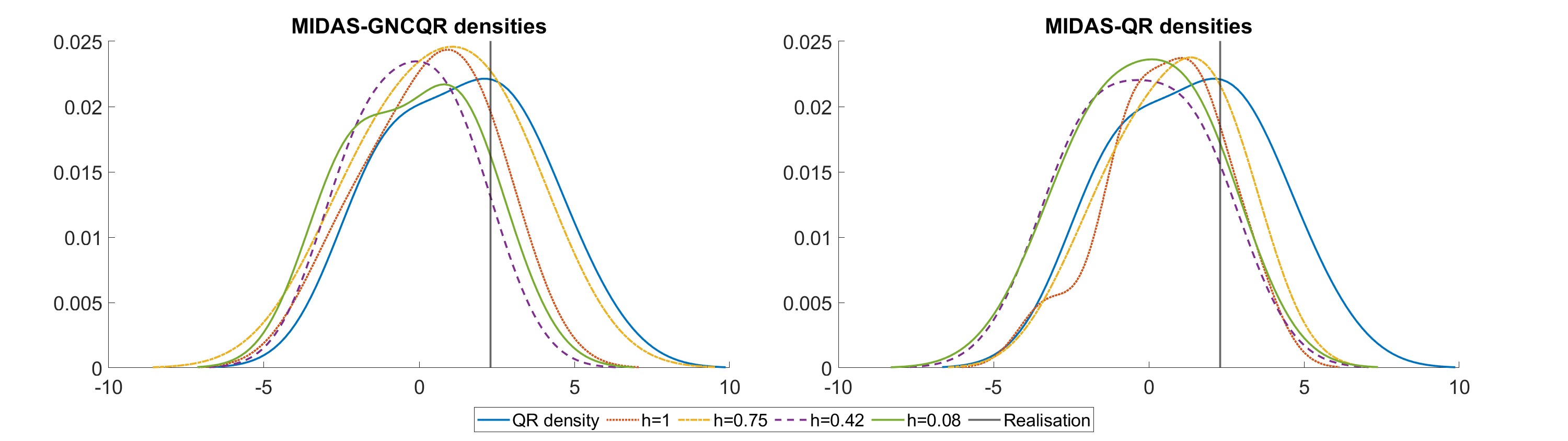}
    \caption{For- and Nowcasted densities of 2008Q2}
    \label{fig:CrisisDensity}
\end{figure}

One key insight of \citet{ferrara2022high}, is that the fitted density can portray large differences at the tails as more timely information becomes available. However, \citet{kohns23hsbqr} and \citet{mitchell2022constructing} show that multimodality emerges in the fitted densities right before crises episodes. To this end we will compare the fitted densities of the MIDAS-QR and MIDAS-GNCQR to see how the fitted densities compare in 2008Q2. In particular, we are interested if whether more timely information gives rise to multimodality. These densities can be found in figure (\ref{fig:CrisisDensity}), which shows 1 forecasted ($h=1$), and 3 nowcasted ($h=0.75$, $h=0.42$, and $h=0.08$) densities. 

The figure reveals several features. First, as more information becomes available, both MIDAS-QR and MIDAS-GNCQR initially nowcast wider distributions, as seen in their respective $h=0.75$ densities, compared to their $h=1$ density. This captures the increased uncertainty that was prevalent in the economy. Second, as more information becomes available the MIDAS estimators move towards the left, i.e. they nowcast lower growth distribution than the simple QR forecast. Third, both MIDAS estimators show some form of multimodality, but they do so at different horizons: the MIDAS-QR shows some multimodality for $h=1$, while the MIDAS-GNCQR does so for $h=0.08$. It is worth noting that for the MIDAS-GNCQR, this multimodality in the GDP distribution emerges as more recent data becomes available, which is not the case for the MIDAS-QR. Note, that with the MIDAS-GNCQR, since there is structure imposed on the quantile dimension, this increased tendency towards multimodality is further evidence to the findings of \citet{mitchell2022constructing} and \citet{kohns23hsbqr}.

\section{Conclusion}
In this paper we extend the MIDAS-QR framework by introducing 2-dimensional structure on the high frequency lag coefficients. Specifically, we approximate the effect of high-frequency lags across the lag dimension with an Almon polynomial as well as constrain the variability across quantiles given a specific lag.

We find that introducing high-frequency data into the estimation framework improves Growth-at-Risk estimates. We also show that the density estimates can be further improved by introducing constraints that regulate the structure across the quantiles. Our proposed method shrinks away quantile variation that does not lead to better out-of-sample fit. In this way we are able to identify high frequency lag profiles that are smooth across the different quantiles. This ability to identify the high frequency variables that have quantile variation, rather than assuming all high frequency variables have quantile variation, lead to better forecast and nowcast performance. This improvement holds across the whole density, as evidenced from alternate weighting schemes considered.

This superior performance of the MIDAS-GNCQR can be potentially further improved. One avenue is to allow the number of polynomials to vary by quantile. Then, cross-quantile constraints can help guard against overfitting, since the different number of polynomials at different quantiles can lead to excess quantile variation given a specific lag. Our findings make a strong argument for 2-dimensional structure in introducing such quantile specific polynomial numbers.

Another avenue for potential improvement is to introduce more high-frequency variables. To address the problems of high-dimensionality in the data, one can introduce shrinkage to select the variables that lead to improved model fit. In the case of high dimensional variables this will require grouped shrinkage profiles so that the polynomials of these variables are jointly shrunk to zero \citep{kohns2022flexible}. Note that this would also need to be implemented in a quantile specific way, so that quantile specific sparsity can be identified.

Finally, one can also extend the hyperparameter regulating the cross-quantile variation to be quantile or variable specific. This can provide further improvements in estimation as parameter specific hyperparameters have shown to offer better performance \citep{zou2006adaptive}. This would necessitate a better hyperparameter tuning procedure as the simple grid-search would become computationally infeasible as the number of hyperparameters increase.

\pagebreak

\bibliographystyle{chicago}
\bibliography{main.bbl}

\pagebreak


\appendix 
\section{Appendix}

\begin{figure}[h!]
    \centering
    \includegraphics[width=\textwidth]{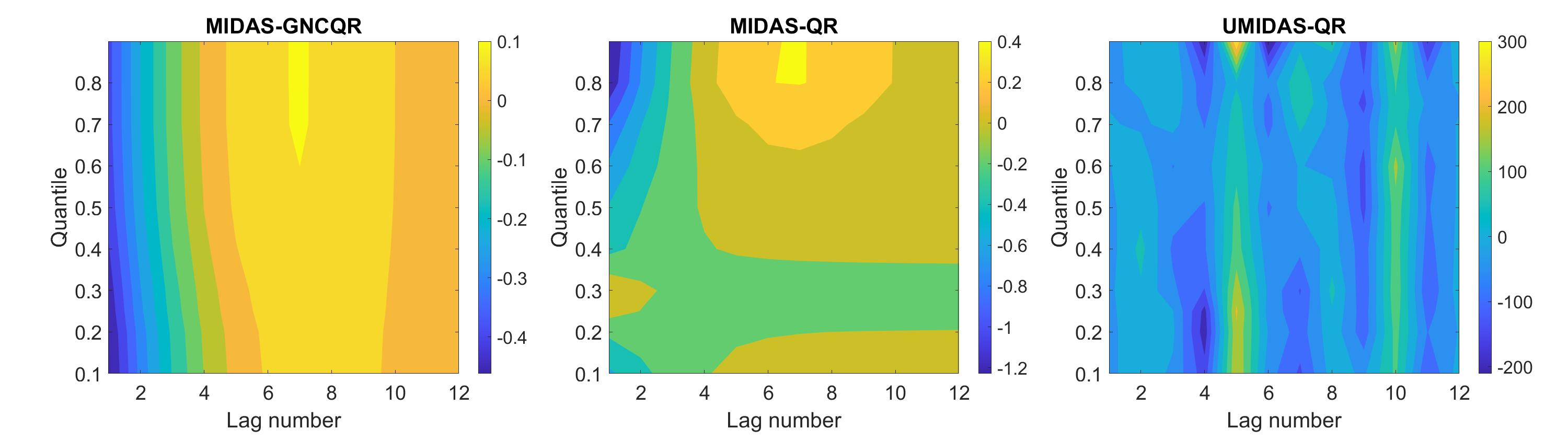}
    \caption{NFCI for h=0.17}
\end{figure}

\begin{figure}[h!]
    \centering
    \includegraphics[width=\textwidth]{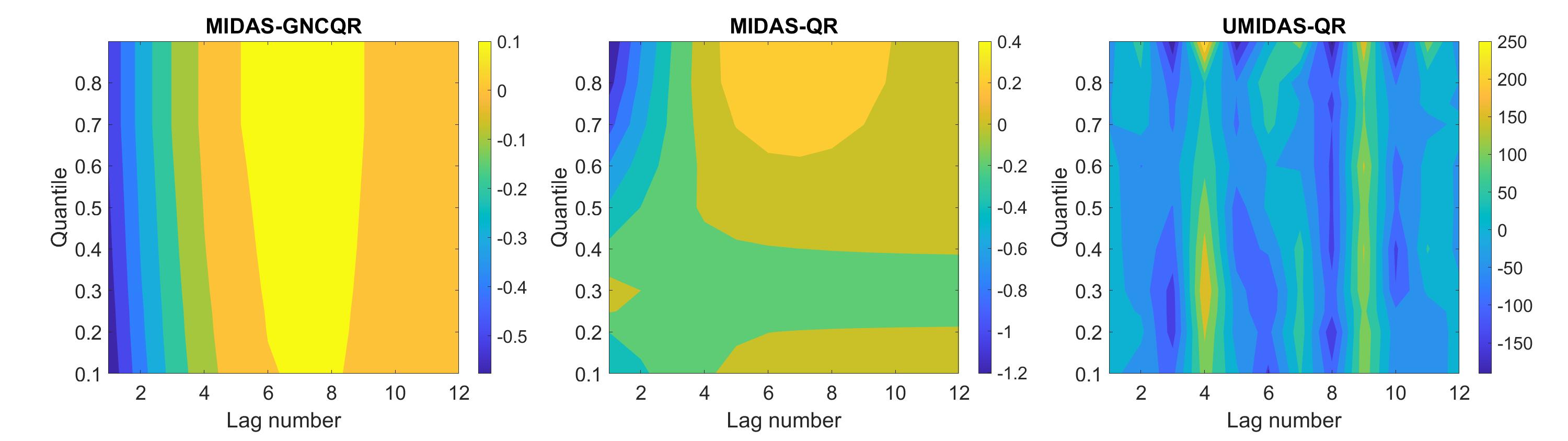}
    \caption{NFCI for h=0.25}
\end{figure}

\begin{figure}[h!]
    \centering
    \includegraphics[width=\textwidth]{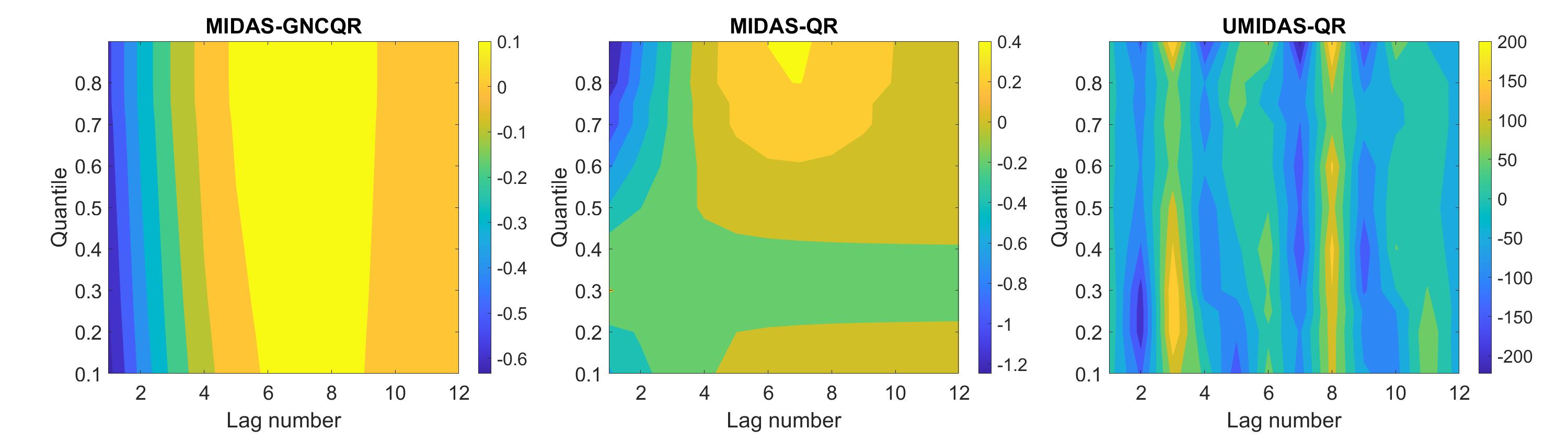}
    \caption{NFCI for h=0.33}
\end{figure}

\begin{figure}[h!]
    \centering
    \includegraphics[width=\textwidth]{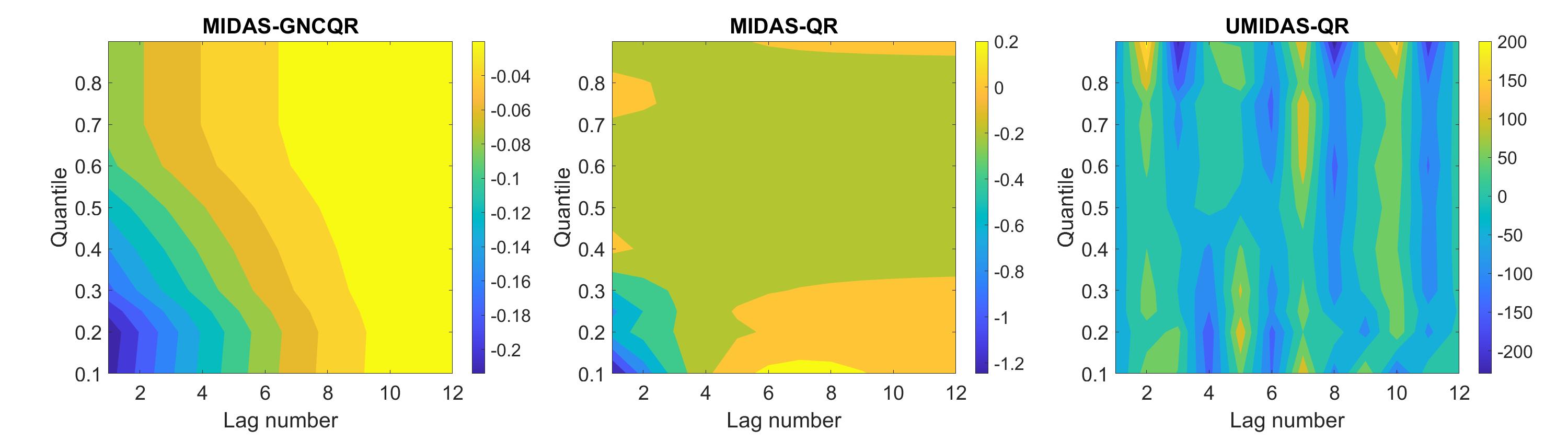}
    \caption{NFCI for h=0.42}
\end{figure}

\begin{figure}[h!]
    \centering
    \includegraphics[width=\textwidth]{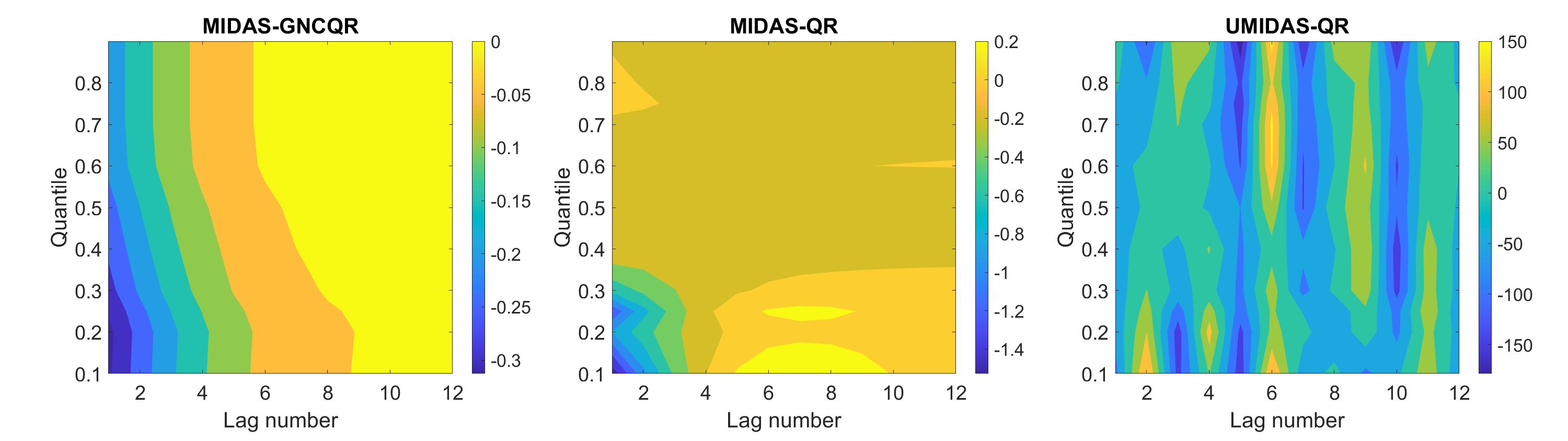}
    \caption{NFCI for h=0.5}
\end{figure}

\begin{figure}[h!]
    \centering
    \includegraphics[width=\textwidth]{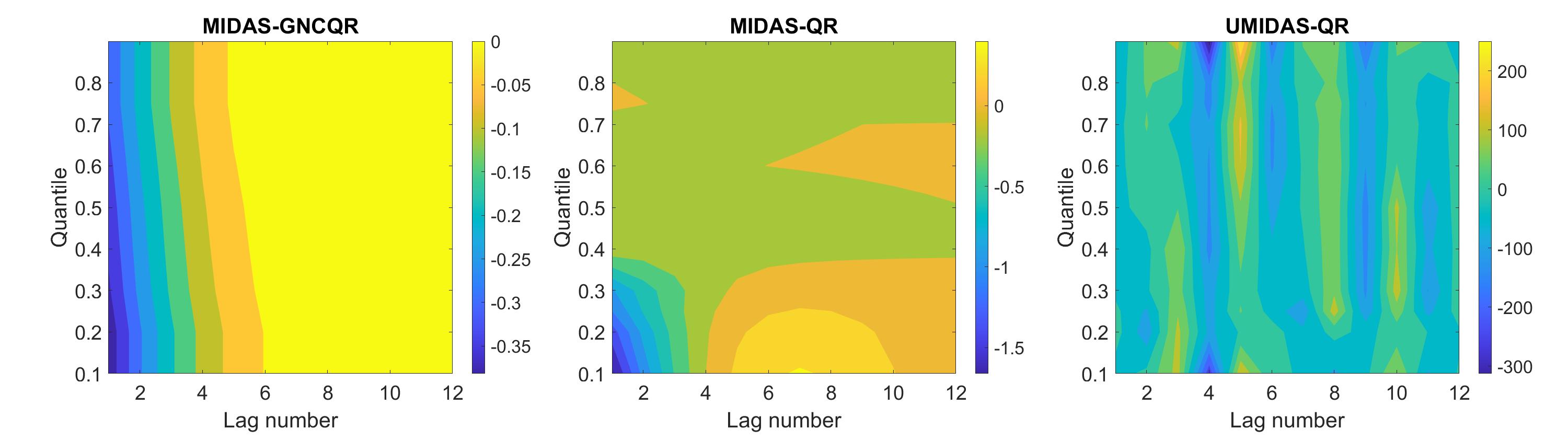}
    \caption{NFCI for h=0.58}
\end{figure}

\begin{figure}[h!]
    \centering
    \includegraphics[width=\textwidth]{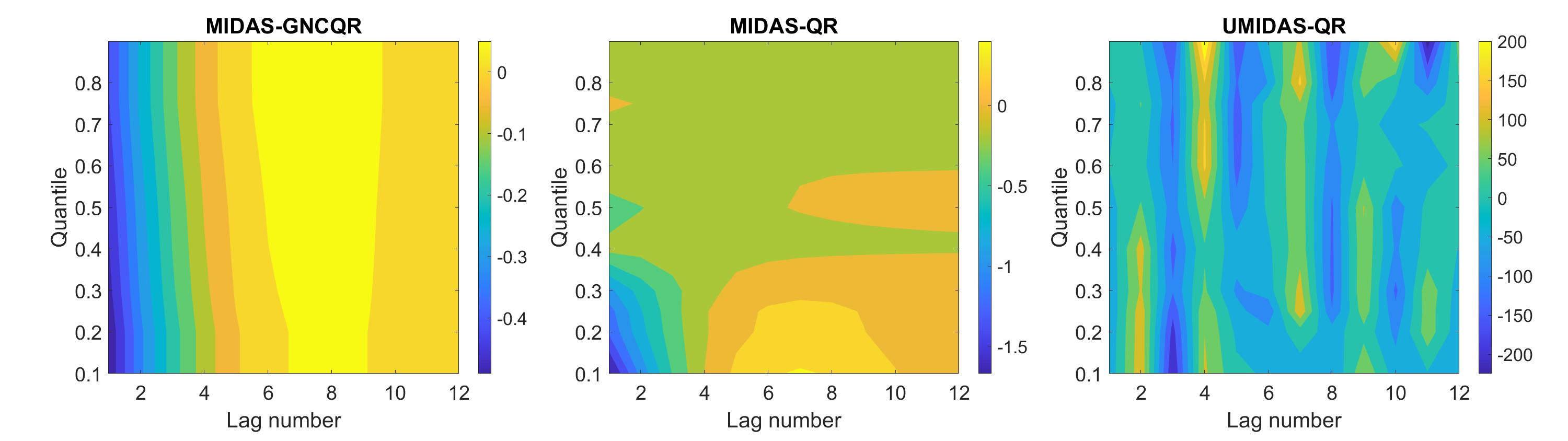}
    \caption{NFCI for h=0.67}
\end{figure}

\begin{figure}[h!]
    \centering
    \includegraphics[width=\textwidth]{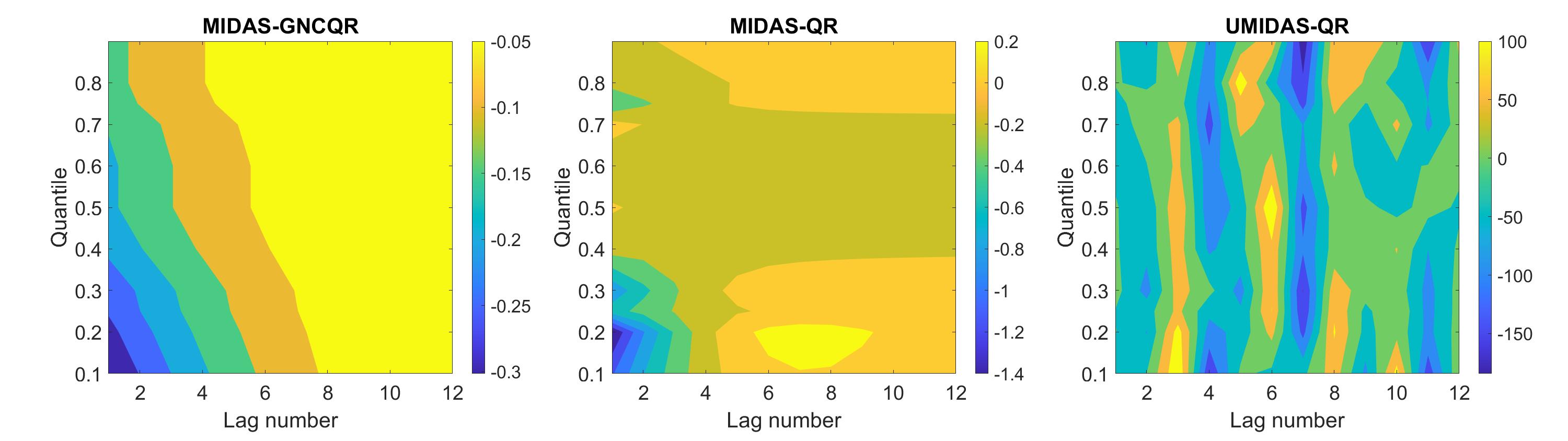}
    \caption{NFCI for h=0.75}
\end{figure}

\begin{figure}[h!]
    \centering
    \includegraphics[width=\textwidth]{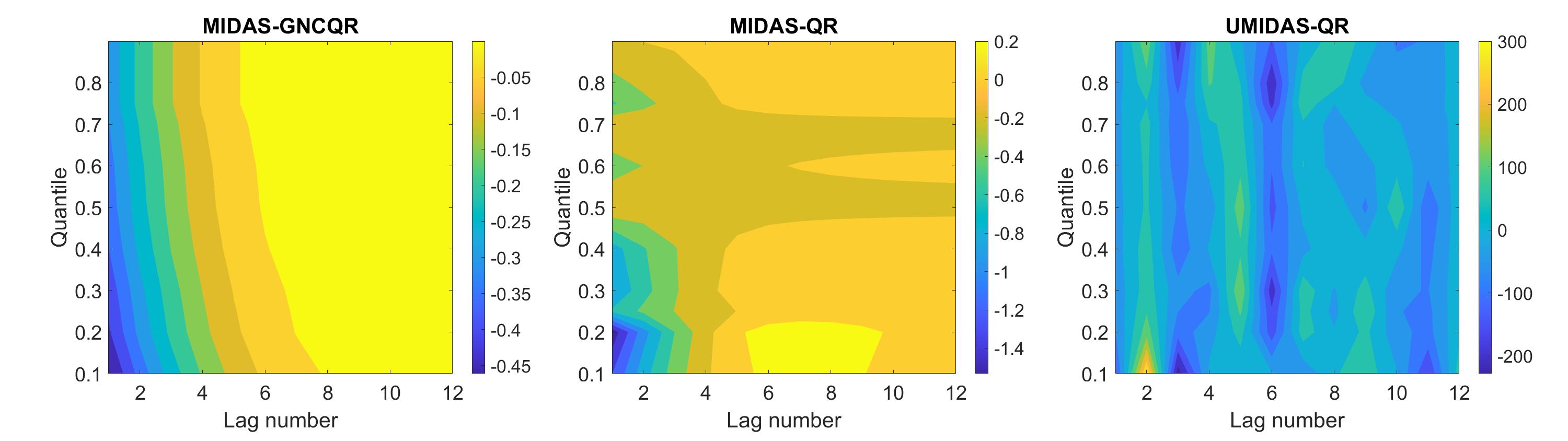}
    \caption{NFCI for h=0.83}
\end{figure}

\begin{figure}[h!]
    \centering
    \includegraphics[width=\textwidth]{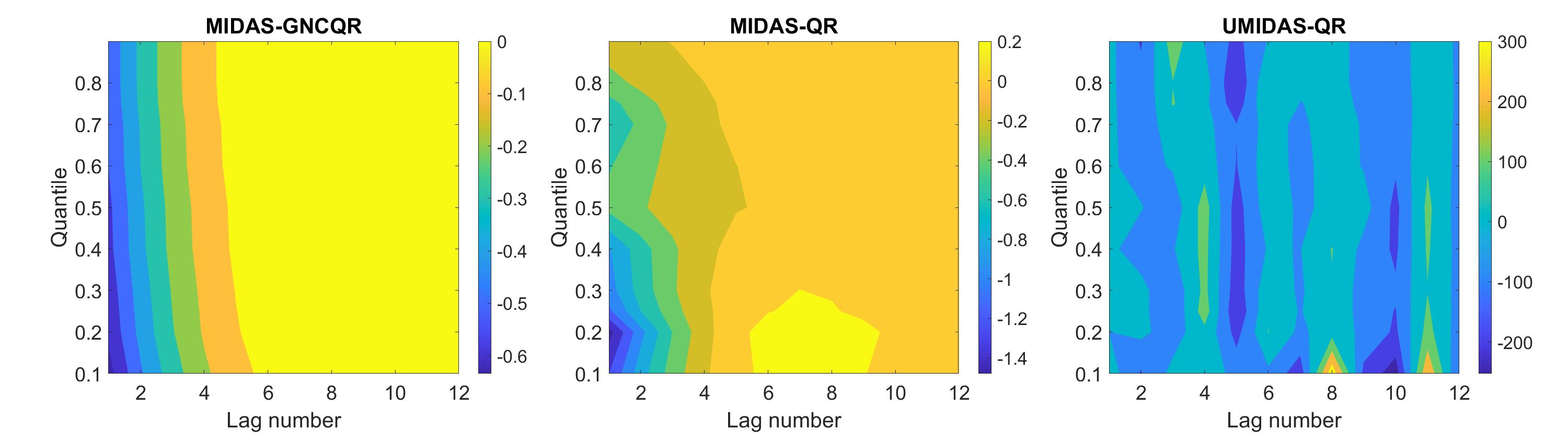}
    \caption{NFCI for h=0.92}
\end{figure}

\begin{figure}[h!]
    \centering
    \includegraphics[width=\textwidth]{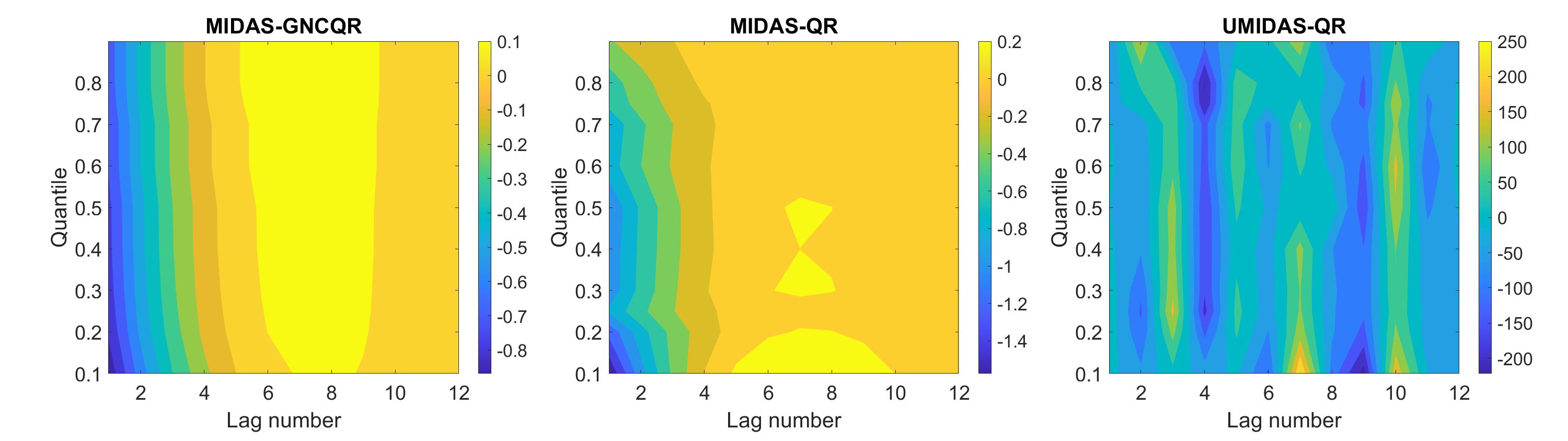}
    \caption{NFCI for h=1}
\end{figure}


\begin{figure}[h!]
    \centering
    \includegraphics[width=\textwidth]{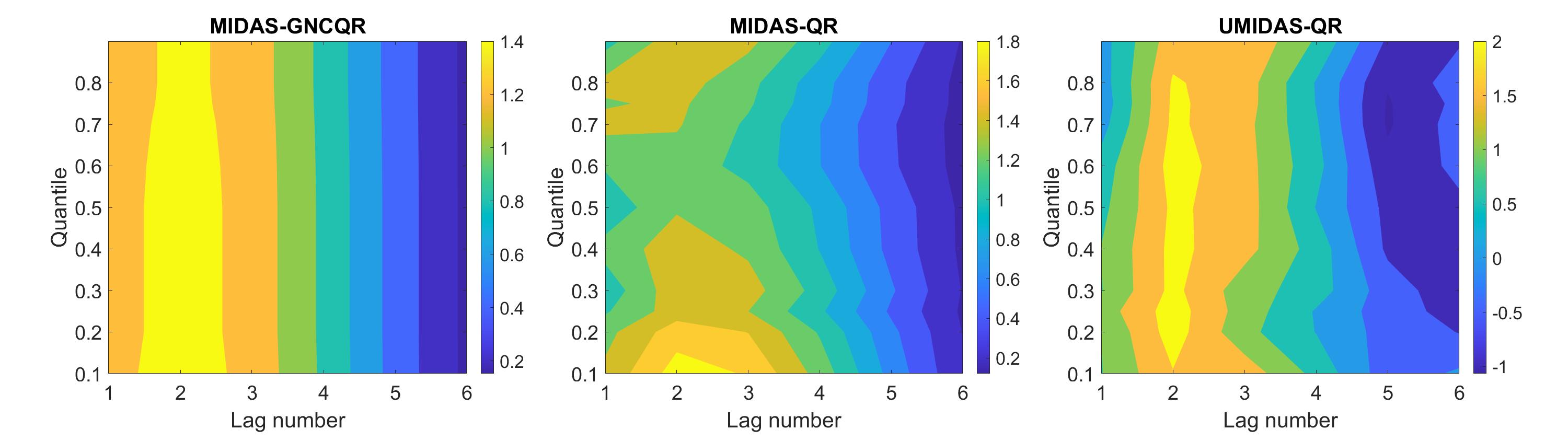}
    \caption{IP for h=0.17}
\end{figure}

\begin{figure}[h!]
    \centering
    \includegraphics[width=\textwidth]{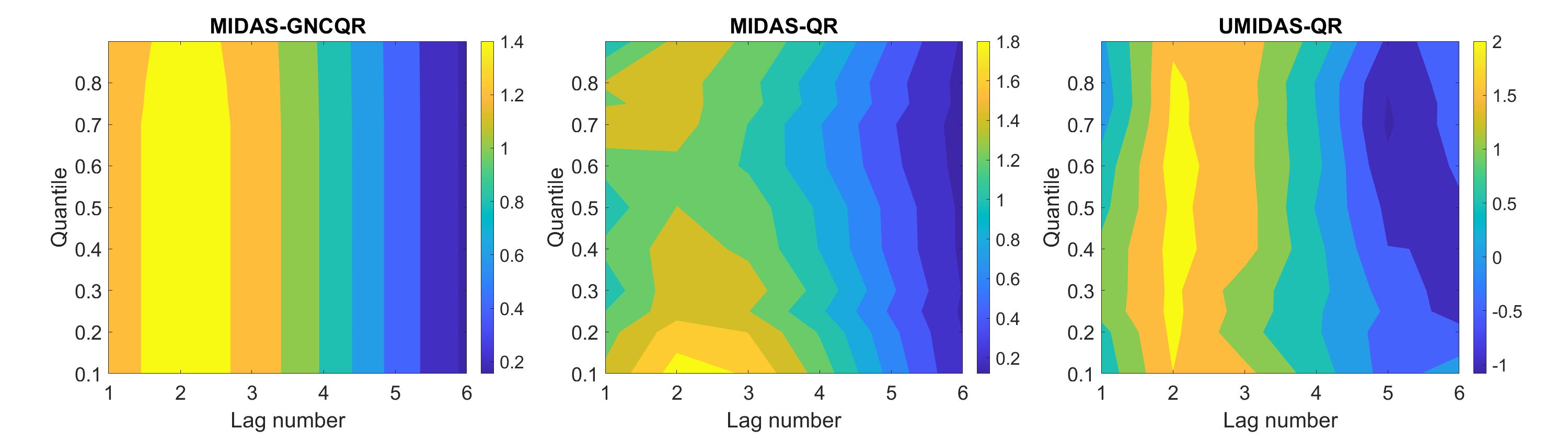}
    \caption{IP for h=0.25}
\end{figure}

\begin{figure}[h!]
    \centering
    \includegraphics[width=\textwidth]{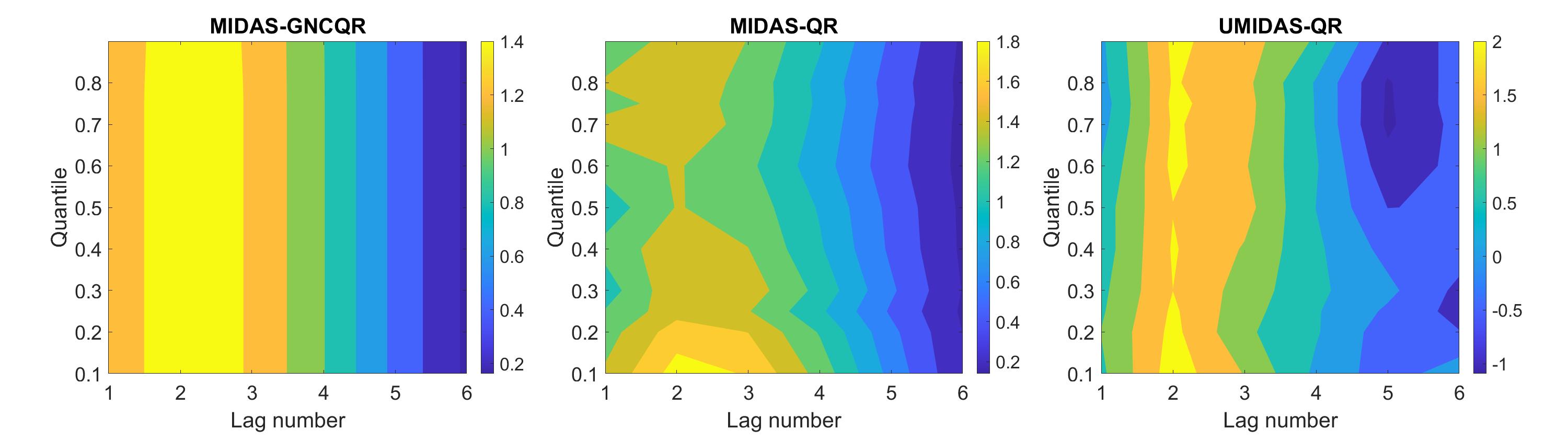}
    \caption{IP for h=0.33}
\end{figure}

\begin{figure}[h!]
    \centering
    \includegraphics[width=\textwidth]{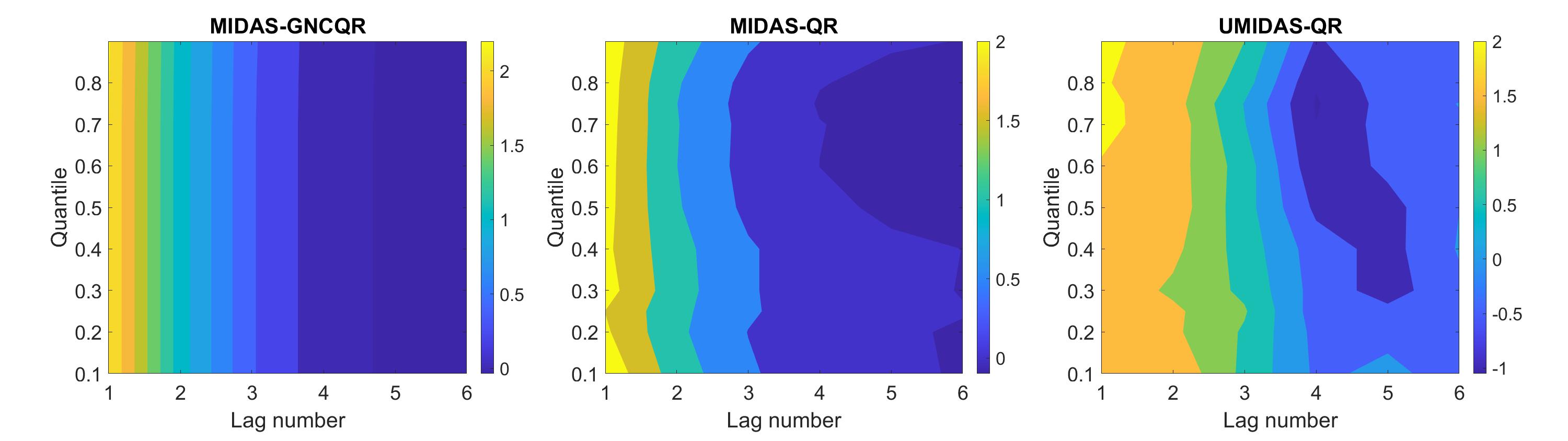}
    \caption{IP for h=0.42}
\end{figure}

\begin{figure}[h!]
    \centering
    \includegraphics[width=\textwidth]{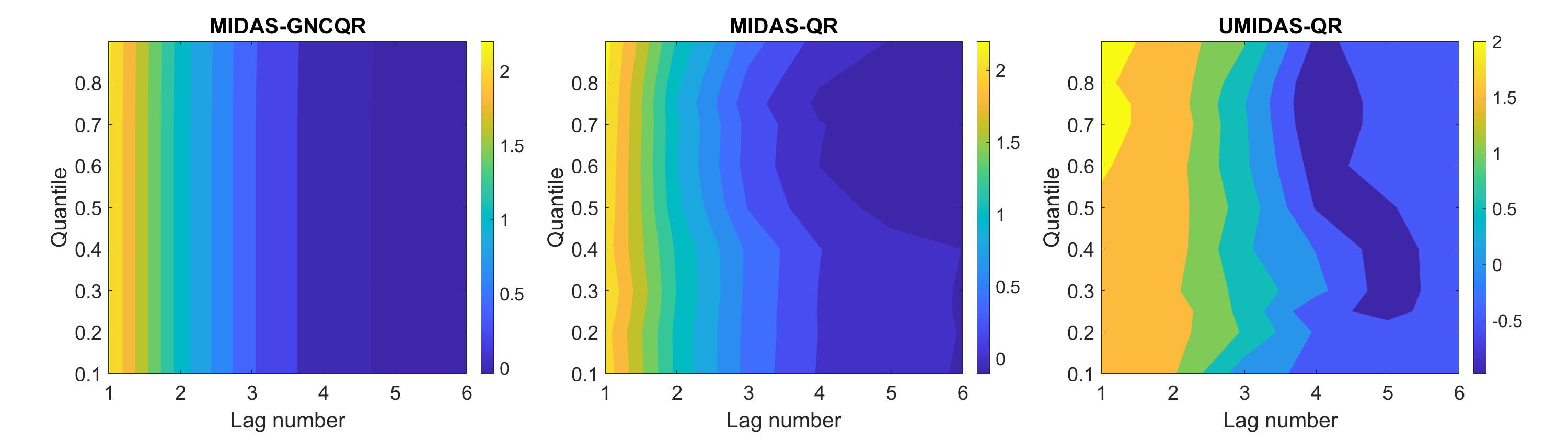}
    \caption{IP for h=0.5}
\end{figure}

\begin{figure}[h!]
    \centering
    \includegraphics[width=\textwidth]{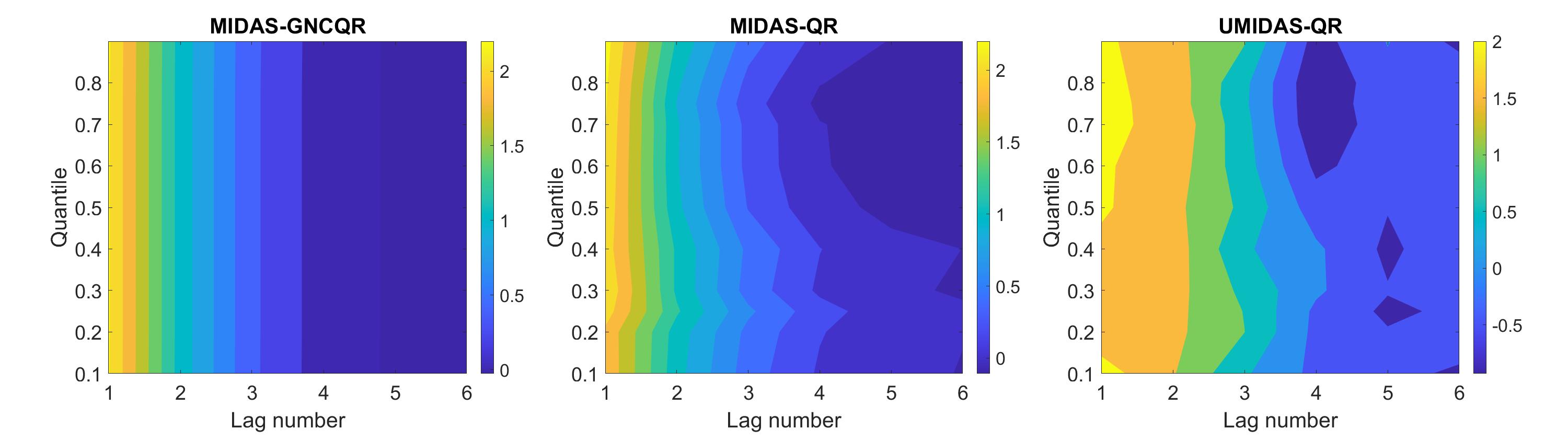}
    \caption{IP for h=0.58}
\end{figure}

\begin{figure}[h!]
    \centering
    \includegraphics[width=\textwidth]{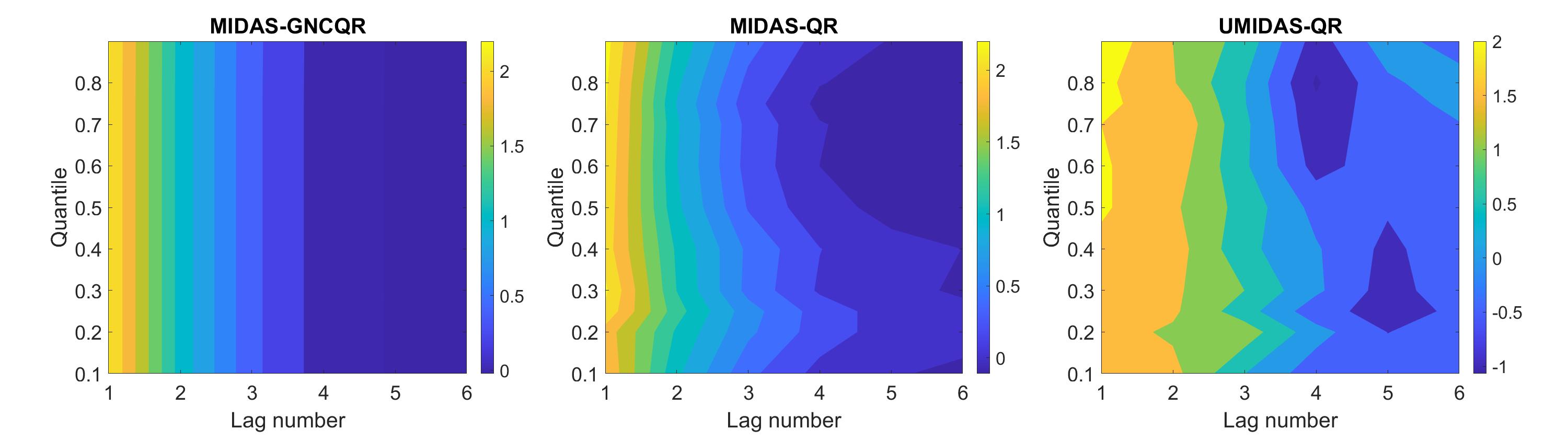}
    \caption{IP for h=0.67}
\end{figure}

\begin{figure}[h!]
    \centering
    \includegraphics[width=\textwidth]{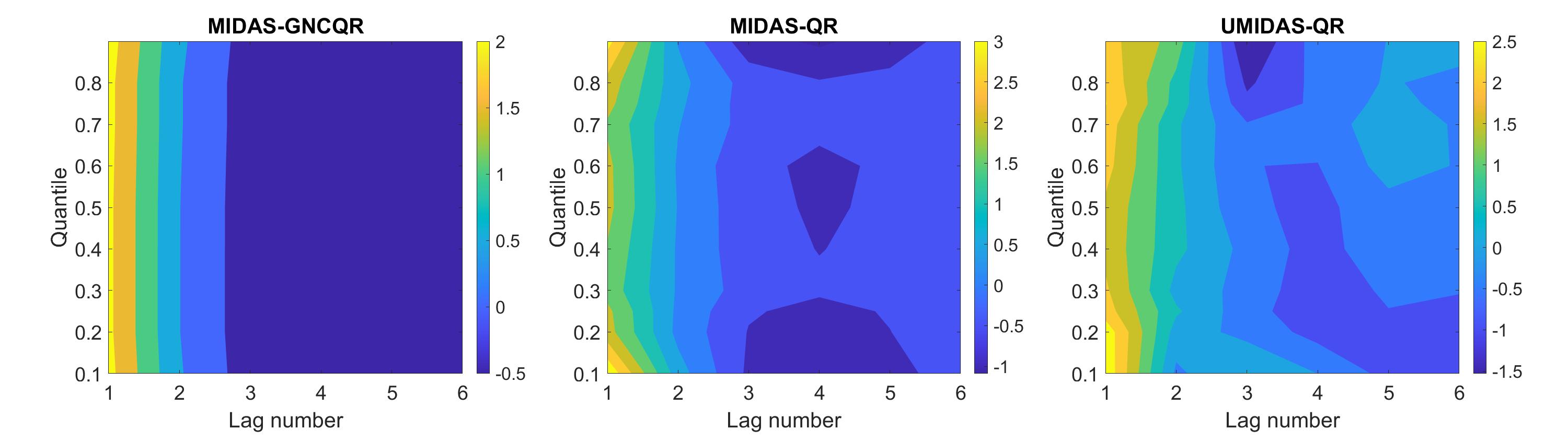}
    \caption{IP for h=0.75}
\end{figure}

\begin{figure}[h!]
    \centering
    \includegraphics[width=\textwidth]{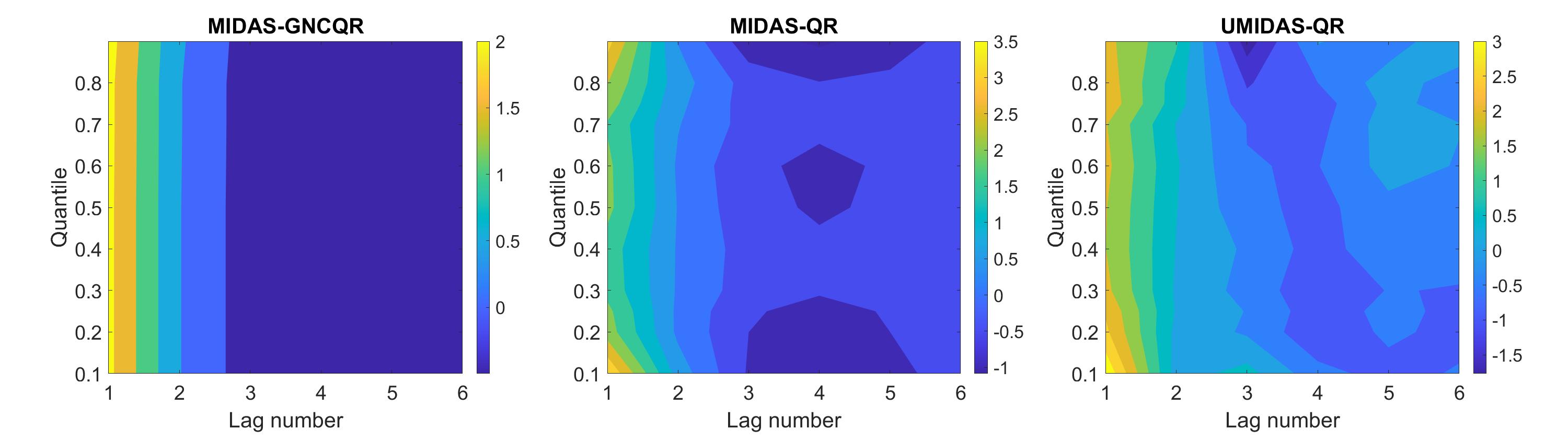}
    \caption{IP for h=0.83}
\end{figure}

\begin{figure}[h!]
    \centering
    \includegraphics[width=\textwidth]{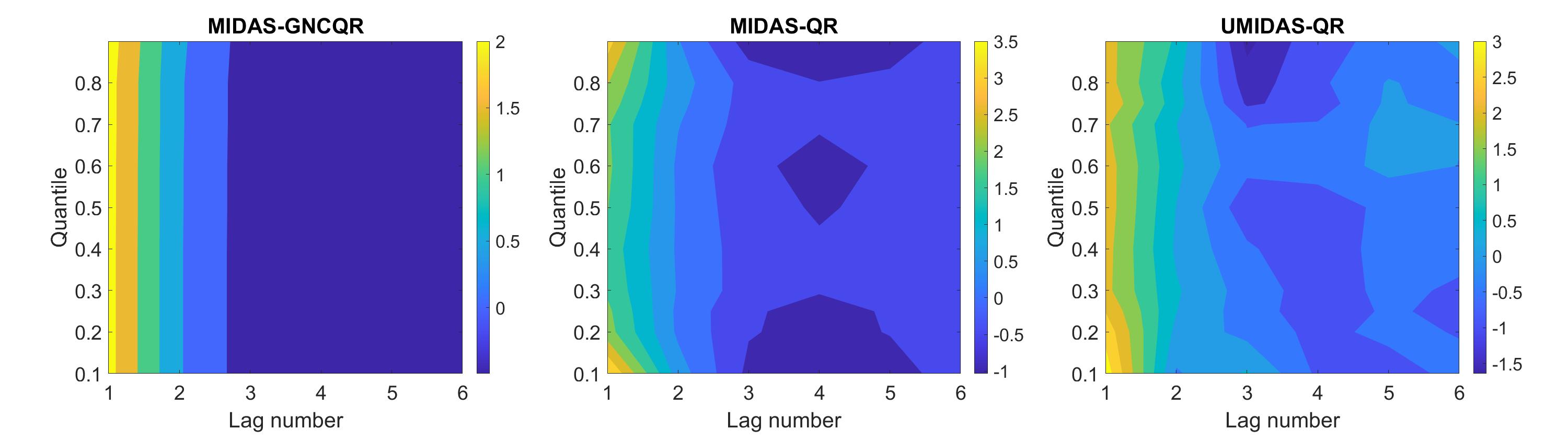}
    \caption{IP for h=0.92}
\end{figure}

\begin{figure}[h!]
    \centering
    \includegraphics[width=\textwidth]{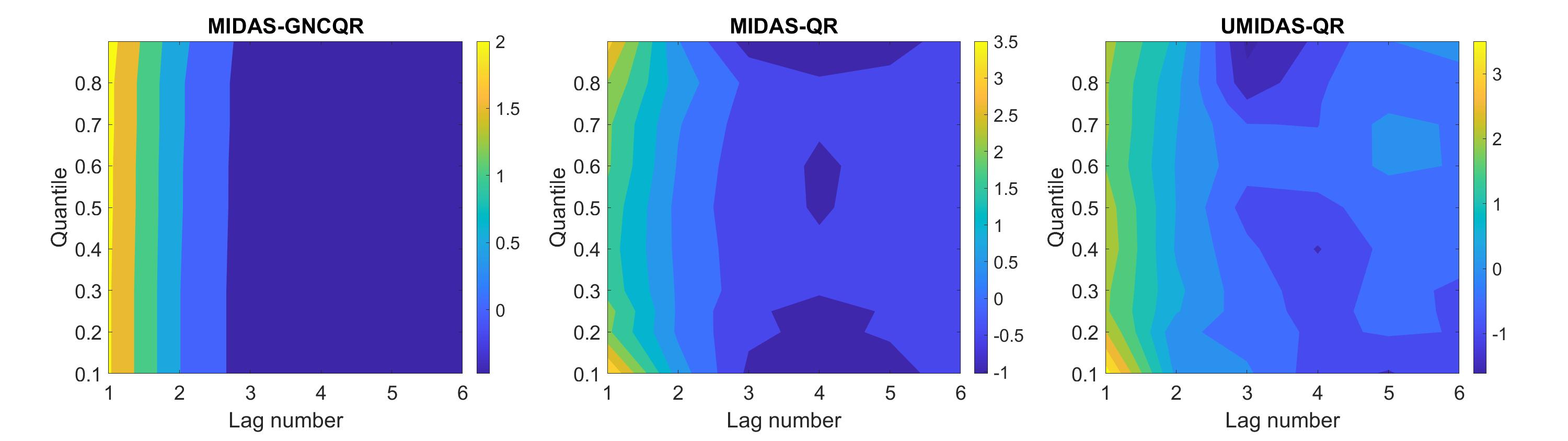}
    \caption{IP for h=1}
\end{figure}


\begin{table}[h!]
\centering
\caption{Nowcast Results: All horizons}
\label{tab:NowRes_App}
\resizebox{\columnwidth}{!}{%
\begin{tabular}{lr|cccc|cccc}
\hline
& & \multicolumn{4}{c|}{Full Sample} & \multicolumn{4}{c}{Pre-COVID}\\
 &  & $\omega_1$ & $\omega_2$ & $\omega_3$ & $\omega_4$& $\omega_1$ & $\omega_2$ & $\omega_3$ & $\omega_4$ \\ \hline \hline
\multicolumn{2}{l|}{h=0.17} &  &  &  &  \\
 & MIDAS-GNCQR & 0.828 & 0.160 & 0.244 & 0.424 & 0.571 & 0.112 & 0.171 & 0.288\\
 & MIDAS-QR & 0.828 & 0.160 & 0.246 & 0.422 & 0.572 & 0.113 & 0.172 & 0.287\\ \hline
\multicolumn{2}{l|}{h=0.25} &  &  &  &  \\
 & MIDAS-GNCQR & 0.814 & 0.158 & 0.243 & 0.413 & 0.567 & 0.112 & 0.169 & 0.286\\
 & MIDAS-QR & 0.829 & 0.160 & 0.246 & 0.424 & {0.572}* & {0.113}* & {0.172}* & 0.287\\ \hline
\multicolumn{2}{l|}{h=0.33} &  &  &  &  \\
 & MIDAS-GNCQR & 0.823 & 0.159 & 0.242 & 0.423 & 0.567 & 0.112 & 0.168 & 0.287\\
 & MIDAS-QR & 0.829 & 0.160 & 0.246 & 0.424 & {0.573}* & {0.113}* & {0.172}* & 0.288\\ \hline
\multicolumn{2}{l|}{h=0.50} &  &  &  &  \\
 & MIDAS-GNCQR & 0.753 & 0.147 & 0.225 & 0.382 & 0.625 & 0.123 & 0.192 & 0.311\\
 & MIDAS-QR & 0.760 & 0.148 & 0.229 & 0.383 & 0.631 & 0.124 & 0.196 & 0.311\\ \hline
\multicolumn{2}{l|}{h=0.58} &  &  &  &  \\
 & MIDAS-GNCQR & 0.748 & 0.146 & 0.222 & 0.380 & 0.624 & 0.123 & 0.189 & 0.312\\
 & MIDAS-QR & 0.757 & 0.147 & 0.228 & 0.382 & 0.628 & 0.124 & 0.194 & 0.310\\ \hline
\multicolumn{2}{l|}{h=0.67} &  &  &  &  \\
 & MIDAS-GNCQR & 0.747 & 0.146 & 0.221 & 0.381 & 0.623 & 0.123 & 0.188 & 0.311\\
 & MIDAS-QR & 0.757 & 0.148 & {0.227}* & 0.382 & 0.628 & 0.124 & {0.193}* & 0.311\\ \hline
\multicolumn{2}{l|}{h=0.83} &  &  &  &  \\
 & MIDAS-GNCQR & 0.917 & 0.176 & 0.282 & 0.459 & 0.640 & 0.125 & 0.192 & 0.324\\
 & MIDAS-QR & 0.930 & 0.178 & 0.286 & 0.466 & 0.645 & 0.126 & 0.195 & 0.324\\ \hline
\multicolumn{2}{l|}{h=0.92} &  &  &  &  \\
 & MIDAS-GNCQR & 0.916 & 0.176 & 0.282 & 0.458 & 0.640 & 0.125 & 0.192 & 0.323\\
 & MIDAS-QR & 0.931 & 0.178 & 0.285 & 0.467 & 0.647 & 0.126 & 0.195 & 0.325\\ \hline
\end{tabular}
}%

{\raggedright \footnotesize Note: Statistically significant differences at the 10\% (*), 5\% (**), and 1\% (*) level are shown. In all cases, MIDAS-GNCQR is the reference estimator for the \citet{diebold1995comparing} test.\par}
\end{table}

\end{document}